\documentstyle[10pt]{article}
\def\l{{\lambda}}
\def\m{{\mu}}
\def\P{{\Phi}}
\def\v{{\varphi}}
\def\a{{\alpha}}
\def\b{{\beta}}
\def\SL{{\cal L}}
\def\nn{ \nonumber }
\def\bq{ \begin{equation} }
\def\eq{ \end{equation} }
\def\ben{ \begin{eqnarray} }
\def\en{ \end{eqnarray} }

\def\on#1#2{\mathop{\vbox{\ialign{##\crcr\noalign{\kern2pt}
 $\scriptstyle{#2}$\crcr\noalign{\kern2pt\nointerlineskip}
\kern-2pt
 $\hfil\displaystyle{#1}\hfil$\crcr}}}\limits}

\def\frac#1#2{{#1\over #2}}
\def\dfrac#1#2{{\displaystyle{#1\over#2}}}

\begin{document}
\title{Stationary problems for equation of the KdV type
and dynamical $r$-matrices}
\author{P.P. Kulish
\thanks{On leave of absence from,
 St.Petersburg's Branch of the Steklov Mathematical
Institute
of Russian Academy of Sciences, 191 011,
 St.Petersburg,  Russia}\\
{\small\it Department of Theoretical Physics and IFIC,
Valencia University,}\\
{\small\it E-46100, Brjassot (Valencia), Spain}\\
S. Rauch-Wojciechowski\\
{\small\it Department of
Mathematics,  Linkoping University,
S-58183,  Linkoping,  Sweden}
\\
A.V. Tsiganov\\
{\small\it
  Department of Mathematical and Computational Physics,
 Institute of Physics,}\\
{\small\it
 St.Petersburg University,
 198 904,  St.Petersburg,  Russia}
}
\maketitle

\begin{abstract}
We study a quite general family of dynamical $r$-matrices for
an auxiliary loop algebra ${\cal L}({su(2)})$ related
 to restricted flows
for equations of the  KdV type. This underlying $r$-matrix structure
allows to reconstruct  Lax representations and to find variables
 of separation
for a wide set of the integrable natural Hamiltonian systems.
As an example,  we discuss the Henon-Heiles system and a
  quartic system of two degrees of freedom in detail.
\end{abstract}


\newpage

\renewcommand
{\theequation}{\arabic{section}.\arabic{equation}}
\newtheorem{theorem}{Theorem}
\newtheorem{prop}{Proposition}
\newtheorem{cor}{Corollary}
\newtheorem{lemma}{Lemma}
\newcommand{\PB}{\stackrel{\textstyle\otimes}{,}}

\section{Introduction}
 The main aim of this paper is to study the $r$-matrix
formulation for mechanical systems
embedded as restricted flows into KdV type equations
and to investigate its connection with Lax representations
 and with the method of separation of variables.
 Lax representations are essentual for
 solving these mechanical systems either
 by linearization on the
 Jacobian of  determinant curve of the Lax matrix
\cite{dnm76}
 or by determining variables of separation
through the functional Bethe ansatz \cite{skl86,skl92}.

Restricted flows are easiest understood as
stationary flows of soliton hierarchies
with sources \cite{arw93}. Many well known integrable
mechanical systems, such as Henon-Heiles, Garnier, Neumann
(and many other) can be embedded into the KdV and
other soliton hierarchies as restricted flows
\cite{arw92a,arw92}.

The main advantage of such embedding is that important
structure elements for restricted flows, such as
Lax representation, bi-Hamiltonian formulation, Newton
pa\-ra\-met\-rization \cite{rw92} can be systematically
derived from the underlying soliton hierarchies.
But it is not the case, as yet, for the $r$-matrix
formulation of these mechanical systems.
There is no direct way of transporting the well
known $r$-matrix formulation for soliton hierarchies
\cite{ft87} to their stationary and restricted flows.
The underlying $r$-matrix structure for finite-dimensional
systems appears to be more complicated then for integrable
PDE's. A new type of, so called, dynamical $r$-matrices
depending both on the spectral parameter and on
dynamical variables has to be considered. The first
example of such $r$-matrix has been found for the
Calogero-Moser system \cite{skl93} and then
for the parabolic and elliptic separable potentials
\cite{eekt94,krw94,ts94}.

It appears that the parabolic and elliptic
 $r$-matrices  belong to
a quite general family of dynamical $r$-matrices  which we
introduce in this paper through a convenient ansatz which
generalizes our $r$-matrix from \cite{krw94,ts94}.
We study the general algebraic properties of these
$r$-matrices
and derive Lax representations through proper
specializations. These specializations are guided by the
known integrable cases of natural Hamiltonian systems
related to restricted flows for equations
of the  KdV type \cite{arw93,krw94}.
Further we use these Lax representations for finding
variables of separation an explain how starting
with variables of separation these Lax representations
be reconstructed by using the Sklyanin approach
 \cite{skl86,skl92}.

The paper is organized as follows.

 In  section 2 we consider  additive algebraic deformations
 of  linear $r$-matrix algebras on an auxiliary loop algebra
 ${\sf g}=\SL({su(2)})$. Kernels of these $r$-matrices
 have  either
 rational, trigonometric and elliptic dependence on  the
 spectral parameter \cite{ft87,fld94,rs87}.
 We consider deformations  $L(\l)=L_0(\l)+F(\l)
 \in \SL(su(2))$ of a certain
 Lax matrix $L_0(\l) \in \SL(su(2))$, which
 satisfies the standard $r$-matrix brackets.
These deformed Lax
 matrices obey a dynamical $rs$-matrix algebra.
 {}From dynamical point of view the deformed
 Lax matrices correspond to systems with  additively shifted
 integrals of motion
 $$I^{new}_k=I^{old}_k + f_k.$$

Further for $r$-matrices of the $XXX$ and $XXZ$
 type we consider  projection of deformations $F(\l)$
 on
 finite-dimensional Poisson (ad$^*_R$-invariant) subspaces
 $\SL_{M,N}(su(2))$. Their Lax matrices $L_{MN}(\l)$ obey
 the $rs$-matrix brackets as well.
 At the end of section 2 we construct  Lax pairs from the
 $rs$-matrix algebra and prove that these Lax pairs are
 related to
 stationary flows for equations of the KdV type.

 In section 3 we express  matrix elements of the matrices
 $L_0(\l)$ and $L_{MN}(\l)$, defined on a direct sum of loop
 algebras  $\oplus \SL(su(2))$, through canonically conjugated
 variables of separation and prove that these matrices obey the
 $rs$-matrix brackets. Such matrices describe a geodesic motion
 and a potential motion on a Riemannian manifold with
 complex diagonal meromorphic Riemannian metrics of special type.
 Corresponding potentials are the finite gap potentials on these
 manifolds.  The problem of quantization is discussed briefly.

 In section 4 we consider
 finite-dimensional integrable systems of  natural type
 and investigate three types of  canonical
 transformations of variables prescribed by the linear
 $r$-matrix structure. Two of these are pure coordinate
transformations
 to generalized elliptic
 coordinates and to Jacobi coordinates, while the third one
 provides an example of a
 momentum
 dependent change of variables. We exemplify of our approach
 with the Henon-Heiles system and a
  quartic potential of two degrees of freedom.

 In the concluding remarks we discuss  shortly an
application of our
  constructions to other
  nonlinear integrable PDE's such as the AKNS and the
 sine-Gordon hierarchies.

\section{Deformations of linear $r$-matrix algebra }
\setcounter{equation}{0}
\subsection{Additive deformations}
Let us recall basic algebraic constructions in the $r$-matrix
scheme \cite{rs87,rs88}
for loop algebras, which we shall use in this paper. A loop
algebra ${\sf g}={\SL}({\sf a})$ is defined as follows.
Let $\sf  a$ be a Lie algebra, then ${\SL}({\sf a})$ consists
 of Laurent polynomials $X(\l)=\sum x_j\l^j$ with
coefficients in ${\sf a}$ for which commutator is defined as
$[x\l^i,y\l^j]=[x,y]\l^{i+j}$.
There is an obvious decomposition of ${\sf g}={\SL}({\sf a})$
 into a linear sum of two
subalgebras
\bq
{\sf g}={\sf g}_+{\on{+}{\,.}}{\sf g}_-\,,\qquad
{\sf g}_+={\mathop{\oplus}}_{j\geq 0}\,{\sf a}\l^j\,,
\qquad {\sf g}_-={\mathop{\oplus}\limits}_{j< 0}
\,{\sf a}\l^j\,.
\label{loop}
\eq
Let $P_\pm$ be  projection operator onto ${\sf g}_\pm$
parallel to ${\sf g}_\mp$.
A standard $R$-matrix associated with ${\SL}({\sf a})$
 is defined as
\bq
R=P_+-P_-\,.
\label{R}
\eq
We shall use  tensor form of  $R$-brackets which is
more common in the inverse scattering method
 \cite{ft87,fld94,ks82}.
If the algebra $\sf a$ is identified with its dual
then the standard $R$-matrix (\ref{R}) is  the Hilbert
 transform in ${\SL}({\sf a})$ with the kernel
\bq
r(\l,\m)=\dfrac{t}{\l-\m}\,,
\label{RR}
\eq
where $t\in{\sf a}\otimes{\sf a}$ is a kernel of the
 identity operator in $\sf a
$. The adjoint kernel is $r^t(\l,\m)=-r(\l,\m)$.
 For twisted subalgebras
${\cal G}({\sf a})\in {\SL}({\sf a})$ the corresponding
decomposition (\ref{loop}) yields kernels with
trigonometric and elliptic dependence on  the
spectral parameter \cite{ft87,fld94,rs87}. In the
 quantum inverse scattering method  $r$-matrices with
  rational, trigonometric and elliptic dependence on
 the spectral parameter
are called $r$-matrices of $XXX$, $XXZ$ and $XYZ$
types respectively \cite{ft87,fld94,ks82}.

In this paper we consider the Lie algebra  ${\sf a}=su(2)$ only.
We fix also representation of the algebra $su(2)$
and shall work in the matrix notation, although the
underlying constructions are independent of particular
 matrix representation. In the following we shall consider
 spin one half representations of $su(2)$ or shall
use the direct sum of these representations. This
means that  we
restrict ourselves  to two-dimensional auxiliary
space in which $su(2)$ matrices act.

There  are two important Lie-Poisson brackets
($R$-brackets)
\cite{rs87,rs88,sur93a} which can be consedered
as linear classical limit of the
fundamental commutator relations
in the quantum inverse scattering method \cite{skl89,fld94}.
They are the $r$-bracket
\bq
\{\on{L}{1}(\l),\on{L}{2}(\m)\}=
[r(\l,\m),\on{L}{1}(\l)+\on{L}{2}(\m)\,]\,,
\label{rpoi}
\eq
and the $rs$-bracket
\bq
\{\on{L}{1}(\l),\on{L}{2}(\m)\}=
[r(\l,\m),\on{L}{1}(\l)+\on{L}{2}(\m)\,]+
[s(\l,\m),\on{L}{1}(\l)-\on{L}{2}(\m)\,]\,,
\label{rspoi}
\eq
which is related to the reflection equations
\cite{skl89,ts94}. Here
we use the standard notations
$\on{L}{1}(\l)= L(\l)\otimes I\,,{~}
\on{L}{2}(\m)=I\otimes L(\m)$.
The brackets (\ref{rpoi}--\ref{rspoi}) are
Lie-Poisson brackets, if  the matrices
$r(\l,\m)$ and $s(\l,\m)$
satisfy the modified classical Yang-Baxter equations
\cite{rs87,rs88,sur93a}.

For a traceless $2\times 2$ matrix $L_0(\l)$
(which we shall call Lax matrix)
\bq
L_0(\l)=\sum_{k=1}^3 s_k\,(\l)\cdot \sigma_k
\label{vop}
\eq
which obeys the linear $r$-matrix algebra (\ref{rpoi})
the $r$-matrix is an antisymmetric function of one
argument $r(\l,\m)\equiv r(\l-\m)$ and
$r(\l)=-r(-\l)$. Namely, put
\bq
r(\l)=\sum\limits^3_{k=1} w_k(\l)\cdot
\sigma_k\otimes\sigma_k\,.\label{r}
\eq
The $\sigma_k$ are  Pauli matrices and
coefficients $w_k(\l)$ are  functions of spectral
parameter only.
They are determined by construction of $R$-matrix on loop
algebra ${\SL}({\sf a})$ (\ref{loop}--\ref{R}) \cite{rs87,rs88} or
by the requirement that the $r$-matrix obeys classical
Yang-Baxter equations \cite{ft87,fld94,ks82}.

It follows from the algebra (\ref{rpoi}) that
the determinant $d_0(\l)\equiv \det L_0(\l)$
  can be taken as
a generating function of commuting integrals of
motion $I_k$ since
\ben
&&\{d_0(\l),d_0(\m)\}=0\,,\qquad
d_0(\l)=\sum_k I_k\l^k\,,\qquad \l,\m\in{\bf C}
\label{det}\\
{\rm implies}\qquad &&\{I_k,I_j\}=0\,.\nn
\en
 A similar property is valid for the $rs$-algebra
(\ref{rspoi}) too.

Here we consider deformations  $L(\l) \in {\SL}({\sf a})$
 of the Lax matrix $L_0(\l) \in {\SL}({\sf a})$ with
 ${\sf a}=su(2)$. The main stimulus
for considering such deformations has been inspired by the study
 of Lax representation and $r$-matrix formulation for
restricted flows of the KdV and coupled KdV hierarchies
\cite{arw93,krw94}, which describe many physically interesting
 integrable natural Hamiltonian systems such as: Neumann,
Garnier, Henon-Heiles systems and an infinite family of
integrable polynomial potentials \cite{wo85}.
The relevant deformations of $L_0(\l)$ had the form
\bq
L(\l)=\left(\begin{array}{cc}a&b\\G(b,\l)
+c&-a\end{array}\right)\,(\l)\equiv
L_0(\l)+F(b,\l)
\eq
where matrix $L_0(\l)\in{\SL}({\sf a})$
and hence the entries $a(\l)\,,{~}b(\l)$
 and $c(\l)$ are Laurent series satisfying
Poisson brackets (\ref{rpoi}) with the $r$-matrix
of the $XXX$ type. The deformation $G(b,\l)$
had a special form connected to KdV equation
\cite{krw94,ts94}.

Here we find a natural generalization of this result
where the deformed matrix $L(\l)$ belongs
the same  phase space. The additive deformation
 of $L_0(\l)$ has the form
\ben
&&L(\l)=\sum_{k=1}^3\left( s_k(\l)+
\a_k(\l)\cdot\dfrac{f(\l)}{2
\tilde{b}(\l)}\right)\cdot
\sigma_k=L_0(\l)+F(\l)\,,
\label{ddop}\\
{\rm where}{~}{~}&&\tilde{b}(\l)=\sum_{i=1}^3 \a_i(\l) s_i(\l)\,,
\label{btilde}\\
{\rm and}{~}{~}&&
\a_1^2(\l)+\a_2^2(\l)+\a_3^2(\l)=0\,.\label{restr}
\en
Here functions $\a_j(\l)$ and the arbitrary
 function $f(\l)$ are complex-valued functions
of the spectral parameter $\l$ only.
The condition (\ref{restr}) guarantes  that
determinant of $L(\l)$ has the additive property
$$d(\l)=d_0\,(\l)+f(\l)\,,$$
which means that integrals of the motion $d(\l)=\sum
I_k\l^k$ of the deformed integrable systems differ
by some constants
$$I^{new}_k=I^{old}_k + f_k,\qquad f_k \in {\bf C}\,.$$

The requirement that the deformed Lax matrix $L(\l)$
satisfy the $rs$-bracket (\ref{rspoi}) with the same
$r$-matrix $r(\l-\m)$ yields four algebraic
equations for the unknown functions $\a_k(\l)\,,{~}k=1,2,3$
and a certain function $g(\l,\m)$.

\begin{theorem} The deformed Lax matrix (\ref{ddop}--\ref{restr})
satisfies the linear $rs$-algebra (\ref{rspoi}), if
\bq
w_j(\l,\m)\a_j(\m)\a_i(\l)-
w_i(\l,\m)\a_i(\m)\a_j(\l)=
g(\l,\m)\cdot\a_k(\l)\,,
\label{restr1}
\eq
where $(j,i,k)$ are cyclic permutations
of indices $(1,2,3)$ and the scale function
$g(\l,\m)$ depends only on the
spectral parameters $\l$ and $\m$.
The corresponding matrix $s(\l,\m)$ is given by
\bq
s(\l,\m)=\sum_{i,j=1}^3 \a_{ij}
(\l,\m)\,\sigma_i\otimes\sigma_j\,,
\label{sg}
\eq
with
\bq
\a_{ij}\,(\l,\m)=
w_i(\l,\m)\cdot g(\l,\m)\cdot\
\dfrac{\a_i(\m)\a_j(\m)f(\m)}
      {\tilde{b}^2(\m)}
-
w_j(\l,\m)\cdot g(\m,\l)
\cdot\dfrac{\a_i(\l)\a_j(\l)f(\l)}
      {\tilde{b}^2(\l)}
\label{alp}
\eq
\label{th1}
\end{theorem}
{\it Proof:} The proof is a direct but lengthy computation.

 Notice that the function $f(\l)$
is quite arbitrary here.
A compact form of the equation (\ref{restr1})
follows from the  ideas of Sklyanin in \cite{skl86}:
\begin{lemma}
The condition (\ref{restr1}) of  Theorem \ref{th1} for the
coefficients $\a_k(\l)$
is equivalent to equation
\bq
\{\tilde{b}(\l),\tilde{b}(\m)\}=
g(\l,\m)\cdot\tilde{b}(\l)-
g(\m,\l)\cdot\tilde{b}(\m)\,.
\label{zerob}
\eq
\label{lem1}
\end{lemma}
{\it Proof:} This lemma is proved by
direct substitution of the
equality (\ref{restr1}) into the
equation (\ref{zerob}) and vice versa.

Let us present some of the best known
\cite{ft87,fld94,ks82}
$r$-matrices and the related solutions
 for $\a_k(\l)$ and $g(\l,\m)$.
 The $r$-matrix of the $XXX$
and $XXZ$ types are
\ben
w_1=w_2\equiv w(\l)=&\dfrac{\eta}{\v(\l)}\qquad
{\rm and}{~} &w_3(\l)=
\dfrac{\eta\v'(\l)}{\v(\l)}\,,\nn\\
\nn\\
{\rm with}{~}{~}{~}{~} \v(\l)=&\l
\qquad&{\rm in{~}the}{~}XXX{~}{\rm case}\,, \label{rxxx}\\
{\rm and}{~}{~}{~}{~} \v(\l)=&\sinh\l
\qquad&{\rm in{~}the}{~}XXZ{~}{\rm case}\,, \label{rxxz}
\en
here $\eta$ is constant and $\v'(\l)$
 denotes derivative with respect to $\l$.

An interesting particular class of solutions
to equations (\ref{restr}--\ref{restr1}) is
distinguished by the conditions
\bq
\a_3=0\,,\qquad \a_1^2+\a_2^2=0\,.
\label{alphs}
\eq
They represent two points on the circle in complex
 plane with the phase difference  equal to $\pi/2$,
for instance $\a_1=1\,,{~}\a_2=i$.
For these solutions the function
 $g(\l,\m)$ is equal to zero ($g(\l,\m)=0)$.
A different example of solutions for $r$-matrix of $XXZ$
 type is
$$\a_1=\cosh\l\,,\qquad \a_2=i\sinh\l
 \,,\qquad \a_3=i\,.$$
Usually one uses the following rational parametrization

 for  $r$-matrix of $XXZ$ type:
\ben
w_1(u-v)=w_2(u-v)&=&\eta\dfrac{2uv}{u^2-v^2}\,,\qquad
w_3(u-v)=\eta\dfrac{u^2+v^2}{u^2-v^2}\,,\nn\\
\label{rxxz2}\\
\v(u)=u-\dfrac{1}{u}\,,&&\quad
{\rm with}\quad u=e^{\l}\,,\quad v=e^{\m}\,,\nn
\en
as, for an example, in the description of the
sine-Gordon equation \cite{ft87}.

An elliptic $r$-matrix of the $XYZ$ type has
\ben
&&
w_1(\l)=\eta\dfrac{\Theta'_{11}}{\Theta_{10}}
\dfrac{\Theta_{10}(\l,k)}{\Theta_{11}(\l,k)}
\,,\qquad
w_2(\l)=\eta\dfrac{\Theta'_{11}}{\Theta_{00}}
\dfrac{\Theta_{00}(\l,k)}{\Theta_{11}(\l,k)}
\,,\nn\\
&&w_3(\l)=\eta\dfrac{\Theta'_{11}}{\Theta_{01}}
\dfrac{\Theta_{01}(\l,k)}{\Theta_{11}(\l,k)}
\,,\nn\\
\label{rxyz}\\
{\rm with}{~}{~}{~}&&\a_1=-1\,;\qquad
\a_2=\dfrac{i}{k}\dfrac{{\rm dn}(\l,k)}
{{\rm cn}(\l,k)}\,,\qquad
\a_3=\dfrac{k'}{k}\dfrac{1}{{\rm cn}(\l,k)}\,.\nn
\en
Here $k$ and $k'$ denote modulus of the Jacobi
elliptic functions
and $\Theta_{ij}(\l,k)$ are the elliptic
theta function in the notation of
\cite{w08}.
 We have to emphasize that
the functions $\tilde{b}(\l)$
(\ref{btilde}) with the coefficients
$\a_k(\l)$ (\ref{rxyz}) and the
 function $g(\l,\m)$ were introduced in \cite{skl86}
for the quadratic $r$-matrix algebra as
solution of the equation (\ref{zerob}).
 The function $g(\l,\m)$ is independent
from second parameter
$\m$ function and it is equal to
$$
g(\l,\m)=\eta\dfrac{\Theta'_{11}}{\Theta_{10}}
\dfrac{\Theta_{10}(\l-K,k)}{\Theta_{11}(\l-K,k)}\,.$$
We recall from \cite{skl86} that, for the purpose of
separation variables, Sklyanin considered a matrix
$\tilde{L}_0(\l)$ which is similar to the
Lax matrix $L_0(\l)$:
\ben
\tilde{L}_0(\l)&=&
\left(\begin{array}{cc}a&b\\ c&-a\end{array}\right)\,(\l)
=U^{-1}(\l)L_0(\l)U(\l)\,,\nn\\
\label{vvop1}\\
{\rm with}\qquad U(\l)&=&
\left(\begin{array}{cc}\Theta_{01}&\Theta_{00}\\
-\Theta_{00}&-\Theta_{01}\end{array}\right)\,
(\xi,\tilde{k})\,,\quad\l=\dfrac{2\xi}{1+k}\,,
\quad\tilde{k}=\dfrac{2\sqrt{k\,}}{1+k}\,,
\nn
\en
Here matrix $L_0(\l)$ obeys the linear $r$-matrix
  algebra (\ref{rpoi})
with the $r$-matrix (\ref{rxyz}).
The similar matrix $\tilde{L}_0(\l)$ satisfies
the $r$-bracket
(\ref{rpoi}) with  $r$-matrix of the form
\ben
\rho(\l,\m)&=&
\on{U^{-1}}{1}(\l) \on{U^{-1}}{2}(\l)\,r(\l,\m)\,
\on{U}{1}(\l)\on{U}{2}(\l)\nn\\
&=&\sum_{j=1}^3
w_j(\l-\m,k)\cdot \sigma_j\otimes\sigma_j+ \nn\\
&+&
\left[\,w_3(\m-K,k)-w_3(\l-K,k)\,\right]
\cdot\sigma_3\otimes\sigma_3+
\nn\\
&+&
i w(\m-K,k)\cdot\sigma_3\otimes\sigma_2
-i w(\l-K,k)\cdot \sigma_2\otimes\sigma_3\,;
\nn\\
\label{rhoxyz}\\
w(\l-\m,k)&\equiv& w_1(\l-\m,k)=w_2(\l-\m,k)=
\eta\dfrac{\Theta'_{11}}{\Theta_{10}}
\dfrac{\Theta_{10}(\l-\m,k)}{\Theta_{11}
(\l-\m,k)}\,,\nn\\
w_3(\l-\m)&=&\eta \dfrac{\Theta'_{11}
(\l-\m,k)}{\Theta_{11}(\l-\m,k)}\,,
\nn
\en
here $K$ is a standard elliptic integral of the module $k$.
Notice that now two first coefficients of the
 matrix $\rho(\l)$
are equal  $w_1(\l)=w_2(\l)$ in same way
as it was in the
$XXX$ and $XXZ$ cases.

It is well known that $r$-brackets (\ref{rpoi}) is the
Lie-Poisson brackets if $r$-matrix obeys the classical
Yang-Baxter equations \cite{ft87,rs87,rs88}. The
Yang-Baxter equation
 is an equation on matrices in the space
${\bf C}^2\otimes{\bf C}^2\otimes{\bf C}^2$.
We use the familiar
tensor notations  $L_j$ for the matrix acting as
 matrix $L$ in the $j$'s factor
 of the product and trivially acting in the remaining
 factors, for example
 $L_1=L\otimes I\otimes I$. In similar way a matrix
$r_{ij}$ is
acting trivially in the third $k$'s factor and it
works as an matrix  $r$ in the
 product of the other $ij$'s factors.
Classical Yang-Baxter equation for pure numerical
$r$-matrices has the form
\bq
[d^\pm_{12}(\l,\m),d^\pm_{13}(\l,\nu)]+
[d^\pm_{12}(\l,\m),d^\pm_{23}(\m,\nu)]+
[d^\pm_{32}(\nu,\m),d^\pm_{13}(\l,\nu)]=0\,.
\eq
Following \cite{skl93,eekt94} we introduce dynamical
Yang-Baxter equations for the
dynamical $rs$-matrix. Let us consider matrices
$d^\pm\equiv r\pm s$ corresponding to
particular class of solutions $\a_k$ (\ref{alphs}).
\begin{theorem}
For the $rs$-algebra (\ref{rspoi}) with  matrices
(\ref{r}--\ref{alp}) the following equations are valid
\ben
&&[d^\pm_{12}(\l,\m),d^\pm_{13}(\l,\nu)]+
[d^\pm_{12}(\l,\m),d^\pm_{23}(\m,\nu)]+
[d^\pm_{32}(\nu,\m),d^\pm_{13}(\l,\nu)]+\nn\\
&&+[L_2(\m),d^\pm_{13}(\l,\nu)]-[L_3(\nu),
d^\pm_{12}(\l,\m)]+\nn\\
&&+[X(\l,\m,\nu),L_2(\m)-L_3(\nu)]=0\,.\label{yb}
\en
The matrix $X(\l,\m,\nu)$ is
\ben
X(\l,\m,\nu)&=&\b(\l,\m,\nu)\cdot\sigma_-
\otimes\sigma_-\otimes\sigma_-\,,  \label{X}\\
&&\nn\\
\b(\l,\m,\nu)&=&
\dfrac{f(\l)b^{-3}(\l)\cdot(\m-\nu)
+f(\m)b^{-3}(\m)\cdot(\nu-\l)+
f(\nu)b^{-3}(\nu)\cdot(\l-\m)}
{(\l-\m)(\m-\nu)(\nu-\l)}\,.\nn
\en
The other two equations are obtained from (\ref{yb})
by cyclic permutation.
\label{theorem3}
\end{theorem}
{\it Proof:} This Theorem also can be proved by a
 straightforward computation.

 Tensor $X(\l,\m,\nu)$ is completely symmetric
tensor with respect to any permutations
of auxiliary spaces.
For the quite arbitrary solutions $\a_k$  to
the equations (\ref{restr}--\ref{restr1}) we can introduce
 an asymmetric tensors $X^{(i,j,k)}(\l,\m,\nu)$.
Then dynamical Yang-Baxter equations
take the general form introduced in \cite{skl93}.

\subsection{Special deformations of the $XXX$ and $XXZ$
 cases and Lax representation}
This section is based on list of formulae collected in
\cite{eekt94,krw94,ts94} and on meaning its in general
$r$-matrix approach to loop algebras.

Our purpose is to apply the $rs$-matrix formulation for
description of natural Hamiltonian systems in order
to solve
them through separation of variables.
For  separating  variables we shall apply the functional
 Bethe ansatz \cite{skl86,skl92}. The Bethe ansatz  has
 an  asymmetric matrix form \cite{fld94,skl92} (variables
 of separation are introduced as zeroes of the entry
$b(\l)$). Therefore we restrict ourselves to a
particular class of deformations obtained by certain
projections of the matrix
\bq
L(\l)=
\left(\begin{array}{cc}a&b\\f(\l)b^{-1}+
c&-a\end{array}\right)\,(\l)=
L_0(\l)+
\left(\begin{array}{cc}0&0\\f(\l)b^{-1}&0
\end{array}\right)\,,
\label{dop}
\eq
where the entries $a(\l)\,,{~}b(\l)$ and $c(\l)$ are
absolutely convergent Laurent series in the $XXX$ case or
absolutely convergent Fourier series in the $XXZ$ case
\ben
a(\l)=&\sum_k a_k\l^k\,,\qquad &{\rm for}{~}XXX{~}
{\rm case,}\nn\\
a(\l)=&\sum_k a_k \exp(k\cdot\l)\,,
\qquad &{\rm for}{~}XXZ{~}{\rm case,}\nn
\en
and similarly for $b(\l)$ and $c(\l)$.

It is
 well known \cite{rs87}  that the standard $R$-bracket on
${\SL}({\sf a})^*$ associated with
 (\ref{loop}--\ref{R}) has a large collection
 of finite-dimensional Poisson (ad$^*_R$-invariant)
subspaces
\bq
{\SL}_{M,N}=\oplus^{N}_{j=-M}\,{\sf a}^*\,\l^j\,,
\qquad{\rm provided}{~}
M\geq 0\,;{~}N\geq-1\,.\label{subsp}
\eq
In other words the subspaces ${\SL}_{M,0}$
and ${\SL}_{1,N}$
 are invariant under the coajoint action of
the subalgebras
${\sf g}_+$ and  ${\sf g}_-$ (\ref{loop}).

Let us introduce the following projection operators
$[\,\cdot\,]_{MN}$
 on these Poisson subspaces
\ben
{[ z ]_{MN}}&=&\left[\sum_{k=-\infty}^{+\infty} z_k\l^k\,
\right]_{MN}\equiv
\sum_{k=-M}^{N} z_k\l^k\,, \nn\\
&&\label{cutmn}\\
{[ z ]_{MN}}&=&\left[\sum_{k=-\infty}^{+\infty}
z_k\exp(k\cdot\l)\,\right]_{MN}\equiv
\sum_{k=-M}^{N} z_k\exp(k\cdot\l)\,.\nn
\en
for the Laurent and for the Fourier series.
In particular in the $XXX$ case
\bq
{[ z ]_+} = \left[\sum_{k=-\infty}^{+\infty}
z_k\l^k\,\right]_+\equiv
\sum_{k=0}^{+\infty} z_k\,\l^k\,,
\label{cut+}
\eq
denotes the  Taylor projection.

We shall define a new additive deformation of
 $L_0(\l)$ as
\bq
L_{MN}(\l)=L_0(\l)+F_{MN}(b,\l)\equiv
\left(\begin{array}{cc}a&b\\G_{MN}(b,\l)+c&
-a\end{array}\right)\,(\l)\,,
\label{dopn}
\eq
where  $G_{MN}(b,\l)$ is a function of the
 spectral parameter $\l$ and
of the entry $b(\l)$. It  reads
\ben
&&G_{MN}(b,\l)=[f(\l)\cdot b^{-1}
(\l)]_{MN},\label{Fmn}\\
&&f(\l)=\sum_{k=-\infty}^\infty f_k\l^k\,,
\quad{\rm or}\quad
f(\l)=\sum_{k=-\infty}^\infty f_k\exp(k \l)
\,.\label{fmn}
\en
For the Taylor projection this deformation takes
the form
\bq
G_N(b,\l)=[f_N(\l)\cdot b^{-1}(\l)]_+,\qquad
{\rm where }{~}f_N(\l)=\sum_{k=0}^N f_k\
\l^k\,.\label{fn}
\eq

The essential feature of this deformation is, in
 comparison with (\ref{ddop}), that
 the determinant of the modified matrix
$L_{MN}(\l)$ is
different from $d_0(\l)+f(\l)$
 $$d(\l)=d_0\,(\l)+b\cdot G_{MN}\neq
d_0\,(\l)+f(\l)$$
 and
therefore integrals of motion  $I_k^{new}$ of the deformed
 system are functionally
different from the undeformed integrals $I_k^{old}$,
and yet the $L_{MN}(\l)$ matrix belongs
to a certain $rs$-algebra.

\begin{theorem}
The matrix $L_{MN}(\l)$ (\ref{dopn})
satisfies the linear $rs$-matrix algebra (\ref{rspoi}),
with the matrix  $s_{MN}(\l,\m)$ given by
\[s(\l,\m)=\a_{MN}\,(\l,\m)\,\sigma_-
\otimes\sigma_-,\quad
\sigma_-=\left(\begin{array}{ll}0&0\\1&0\end{array}
\right).\]
The function $\a_{MN}\,(\l,\m)$ is defined by
\bq
\a_{MN}\,(\l,\m)=
-w(\l-\m)\,\left( [f(\l) b^{-2}(\l)]_{MN}-
[f(\m) b^{-2}(\m)]_{MN}\,\right)\,,\nn\\
\label{alpn}
\eq
where $w(\l)\equiv w_1(\l)=w_2(\l)$.
\label{theorem2}
\end{theorem}
{\it Proof}:
The proof is a
straightforward calculation.

The Theorem  3 has a transparent meaning in the
framework of the general
$r$-matrix approach to  loop algebras.
The Theorem 3 asserts that  the matrix $s(\l,\m
)$ can be restricted to subspaces ${\SL}_{M,N}$
(\ref{subsp}) and subspaces ${\SL}_{M,N}$
is a common Poisson subspace for the $rs$-brackets
 (\ref{rspoi})
 with matrix $s_{MN}=[s(\l,\m)]_{MN}$.
 Since the deformations (\ref{vop})
 are directly connected with the KdV equation
\cite{krw94},
Theorem 3 describes
the relationship between compatible Poisson
 structures of
the KdV equation
and the Virasoro algebra \cite{avm94}.
Matrices $d^\pm=r \pm s_{MN}$ obey dynamical Yang-Baxter
 equations, which
are obtained by
restriction of the equation (\ref{yb}--\ref{X}) to the
 corresponding subspace ${\SL}_{MN}$. For certain
 matrices $L_{MN}(\l)$ and $s_{MN}$ this proposition
has been proved in \cite{eekt94}.

In order to describe Lax representations for  integrable
Hamiltonian systems related to
matrices $L_0(\l)$ and $L_{MN}(\l)$ we
introduce  matrices \cite{rs87,rs88}
$$d_{12}(\l,\m)=r(\l,\m)+s(\l,\m)\,,
\quad{\rm and}\quad
d_{21}(\l,\m)=r(\l,\m)-s(\l,\m).$$
In terms of these matrices the Lie-Poisson
$rs$-brackets reads:
\bq
\{\on{L}{1}(\l),\on{L}{2}(\m)\}=
[d_{12}(\l,\m),{~}\on{L}{1}(\l)]+
[d_{21}(\l,\m),{~}\on{L}{2}(\m)]\,,\label{dd}
\eq
For  constructing  Lax representations  we
shall make use of  the following lemma \cite{bv90}.
\begin{lemma}
If matrix $L(\l)$ obeys the Lie-Poisson algebra (\ref{dd})
then the spectral invariants
\bq
H_{n}(\l)={\rm tr}\,(L^n)\,,\qquad n=1,2,\ldots;
\,,\label{hn}
\eq
are integrals of motion in
involution $\{H_n,H_m\}=0$
 and the  Lax equation are given by
\ben
&&\dot{L}(\m)=\{H_n(\l),L(\m)\}=
[M_n(\l,\m),L(\m)],\\
{\rm with}\qquad&&M_n(\l,\m)=n\,{\rm tr}_1
\left(\,\on{L^{n-1}}{1}(\l)\,d_{21}(\l,\m)
\,\right),{~} n=1,2\ldots\,.
\en
where ${\rm tr}_1$ means trace in the first auxiliary
 space and dot over $L$ means derivative with respect
to time corresponding to the Hamiltonian $H_n(\l)$.
\label{lem3}
\end{lemma}
In two dimensional auxiliary space
$$d(\l)\equiv det L(\l)=-\dfrac12 ({\rm tr}\,(L^2)
=-\dfrac12 H_2(\l)\,.$$

In order to get a Hamiltonian $H$, which does not
depend on the
 spectral parameter, one needs a linear functional
 (a projection) $\P_\l$ which selects,
 for example (in $XXX$ case),
 a coefficient at certain power of $\l$
\bq
H=\dfrac12\P_\l\,[H_2(\l)]=
\P_\l\,[d(\l)]\,.
\label{ham}
\eq
 For instance
it can be defined as a residue of order $m$ at  $\l_0$
\bq
\P_\l\,[z]={\rm Res}^m_{\l_0}\,z(\l)=
\dfrac{1}{(m-1)!}\left.
\frac{d^{m-1}}{d\l^{m-1}}\,\left((\l-\l_0)^m
\,z(\l)\right)
\right|_{\l=\l_0}
\eq
Then the Lax representation for the Lax matrices (\ref{vop}
 or \ref{dop})
with the Hamiltonian (\ref{ham}) has the form
\ben
\dot{L}(\m)&=
&\{H,L(\m)\}=\{\dfrac12\P_\l\,[\,{\rm tr}_1\,
\on{L^2}{1}\,(\l)\,],L(\m)\}
=[M(\m),L(\m)]\,,\nn\\
&&\label{glax}\\
M(\m)&=&\P_\l\,\left[{\rm tr}_1\,
\left(\on{L}{1}\,(\l)\cdot d_{21}(\l,\m)\,\right)\right]
=\P_\l\,\left[\,M(\l,\m)\,\right]\,.\nn
\en
In particular for the initial Lax matrix $L_0$ (\ref{vop}) one
 gets $s=0;{~}{~}d_{21}(\l,\m)=r(\l-\m)$ and then
\ben
\dot{L}_0(\m)&=&\left[M_0(\m),L_0(\m)\right]\,,,\label{vm}\\
M_0(\m)&=&\P_\l\left[\,
2\sum_{k=1}^3 s_k(\l)\cdot w_k(\l-\m)\cdot\sigma_k\,
\right] \,
\en

 In this paper we start with the requirement that
\bq
M_0=\sigma_+\equiv \left(\begin{array}{cc}0&1\\0&0
\end{array}\right)\,.
\label{vam}
\eq
and choose $L_0(\l)$ and $\P_\l$ appropriately.
It is a rather strong restriction for the matrices $L_0$
 (\ref{vop}) and for the corresponding Hamiltonians.
This requirement comes from the study of $rs$-representations
 for natural Hamiltonian systems related to
 stationary flows for equations of KdV type
\cite{krw94,ts94}.

As an immediate consequence of the Lax representation
 (\ref{glax})
with the matrix  $M_0$ (\ref{vam})
we can rewrite the Lax matrix $L_0$ in the form
\bq
L_0(\m)=\left(\begin{array}{cc}-\dfrac{b_x}2&b\\
-\dfrac{b_{xx}}2&\dfrac{b_x}2\end{array}\right),\quad
b_x=\{H_0,b\},
\label{vopkdv}
\eq
where $H_0$ is a Hamiltonian corresponding to $L_0$.
Since the determinant of $L_0(\l)$ is a generating
function of integrals of motion  we see that it obeys
the equation
\bq
-\{H_0\,,d_0(\l)\}=
-d_{0,x}(\l)=\partial^3_x \cdot b=b_{xxx}=0\,. \label{veqmot}
\eq

Now for a fixed triad ($L_0$, $M_0$,
$\P_\l$) we introduce a modified matrix $L(\l)$ (\ref{dop}).
In this case the
matrix $M(\m)$ is constructed according to (\ref{glax}) with
$d_{21}\,(\l,\m)=r-s$ and with the use of the same projector
 $\P_\l$. It reads
\bq
M(\m)=
\left(\begin{array}{cc}0&1\\-u(\m)&0\end{array}\right)\,,
\qquad u(\m)=f(\m)b^{-2}(\m)\,.
\label{dem}
\eq
due to the  linearity of $\P_\l$
\ben
\P_\l\left[\,w_1\cdot
\left(\,-f(\l)\,b^{-1}\,(\l) +c(\l)+
(f(\l) b^{-2}(\l)-f(\m) b^{-2}(\m))\cdot b(\l)\,\right)
\right]=\nn\\
\label{phiprom}\\
=-f(\m) b^{-2}(\m)=-u(\m)\,.\nn
\en
Then from the Lax representation (\ref{glax}) it follows
that the deformed matrix $L(\l)$ has to be equal to
\bq
L(\l)=\left(\begin{array}{cc}-\dfrac{b_x}2&b\\
-b(\l)u(\l)-\dfrac{b_{xx}}2&\dfrac{b_x}2\end{array}
\right),\quad
b_x=\{H,b\}.
\label{dopkdv}
\eq
where the Hamiltonian $H=\P_\l[d(\l)]$ is
obtained by applying the old projector $\P_\l$
to the modified determinant.
The determinant $d(\l)$ of the modified matrix
$L(\l)$ obeys the equation
\bq
-d_x(\l)=(\dfrac14\partial^3_x +u\partial_x
+\dfrac12 u_x\,)\cdot b=B_1[u]\cdot b=0. \label{deqmot}
\eq
Here $B_1[u]$ is the Hamiltonian pencil operator for
 the coupled KdV equation \cite{arw92,arw93}.

For the deformed (according to (\ref{dopn})) matrix
 $L_{MN}(\l)$
the Lax equation is constructed analogously
(\ref{dem}--\ref{dopkdv}) and (\ref{deqmot}).
One has to replace $u(\m)=f(\m)b^{-2}(\m)$ with the potential
$u_{MN}(\m)=[f(\m)b^{-2}(\m)]_{MN}$.

 Equation (\ref{deqmot}) shows  that our deformation is
 connected to the stationary problems for equation
 of the KdV type.
The Hamiltonian and some properties of motion
 for the finite-dimensional integrable systems
connected with stationary flows of the KdV equations
were considered in \cite{al81,al92,almar92}. Lax
representations,
 bi-Hamiltonian
 structures, Newton representations  and some other
 properties
for these systems
 were introducing in \cite{arw92,arw92a,rw92}
through method of
 the restricted flows for nonlinear equations.
Some particular systems were studied in the
$r$-matrix approach
\cite{eekt94,krw94,ts94}. In the next section
we tie these different
  methods
to the Lie algebraic method of constructing matrices $L(\l)$
 on the loop algebras.

In the next section we shall assume that the Lax
matrix $M_0$ has the form
\bq
M_0=\left(\begin{array}{cc}v_0&u_0\\0&-v_0\end{array}\right)\,.
\label{vamcom}
\eq
It introduces certain technical complications
without influence
on the conceptual structure.

\section{Construction of Lax matrices $L(\l)$
for separable systems}
\setcounter{equation}{0}
\subsection{Separation of variables}
The main problem for a given integrable Hamiltonian
system with a
complete set of functionally
independent and commuting integrals of
motion $H_1\,,\ldots,H_n$
is to determine its solutions.
The construction in the Liouville theorem has
a local character
and there is no general way to describe solutions
 globally unless
 some additional information about the  system
{ is known},
such as a complete, spectral parameter
dependent Lax representation or variables of
separation.

Here we study the  Lax matrices satisfying the
linear $rs$-algebra
 (\ref{rpoi}--\ref{rspoi}) and method of
 separation of variables.
By separation of variables we mean here
 construction of $n$ pairs of canonically conjugate
 variables
\bq
\{u_j,u_k\}=\{v_j,v_k\}=0\,,\qquad
\{v_j,u_k\}=\delta_{jk}\,,\qquad j,k=1,\ldots,n\,
\eq
and functions $\Psi_k$ such that
\bq
\Psi_k(u_k,v_k,H_1,H_2,\ldots,H_n)=0\,,\qquad k=1,\ldots,n\,
\eq
where $H_k$ are integrals of motion
in involution \cite{skl92}.
Note that the canonical transformation
 from the original variables
$(x_k,\,p_k),{~}k=1,\ldots,n$ to variables of separation
$(u_k,\,v_k),{~}k=1,\ldots,n$ may not necessarily
 be a pure coordinate
change. It can involve both coordinates and momenta.

For the Lax matrices
$L(\l)$ satisfying $r$-matrix algebra with
the $r$-matrix of the $XXX$ type
the problem of  determining  variables of separation
 has been essentially solved by Sklyanin
(see Theorem 4 below).

Here we provide a solution of the converse problem how
 to construct Lax matrices $L(\l)$ satisfying the
$rs$-matrix algebra for an integrable system given
in terms of variables of separation. We express first
$L(\l)$ in terms of variables of separation
$(u_k,\,v_k),{~}k=1,\ldots,n$ and then we shall use the
 canonical transformations
naturally determining from the $r$-matrix structure
to the "physical" variables $(x_k,\,p_k),{~}k=1,\ldots,n$.

\begin{theorem}[Sklyanin \cite{skl86,skl92}]
\par\noindent
Coordinates $u_k,\,v_k,{~}{~}k=1,\ldots,n\,$ defined as
\bq
b(\l=u_k)=0\qquad v_k=a(\l=u_k)
\label{sepq}
\eq
are  canonically conjugate variables of separation,
if the matrix $L(\l)$ obeys the $r$-matrix
algebra with
$r$-matrices (\ref{rxxx}) or (\ref{rxxz}).
Separation equations are
\bq
v_k^2=-d(u_k)\,, \label{seq}
\eq
where $d(\l)\equiv {\rm det}\, L(\l)$.
\label{th2}
\end{theorem}
{\it Proof:}
We repeat the Sklyanin proof for the $XXX$
and $XXZ$ cases,
because this technique  will be used  later.

Let us list  commutation relations for $b(\m)$
and $a(\l)$,
\ben
\{b(\l),b(\m)\}&=&\{a(\l),a(\m)\}=0,
\label{uuvv}\\
\{a(\l),b(\m)\}&=&
w_3(\l-\m)b(\m)-w(\l-\m)b(\l)\,,
\label{vuvu}\en
where we denote that $w(\l-\m)\equiv
w_1(\l-\m)\equiv
w_2(\l-\m)$.
The equalities $\{u_i,u_j\}=0$ follow from (\ref{uuvv}).
To derive the equality $\{u_i,v_j\}=\eta\delta_{ij}$
we substitute
$u=u_j$ in (\ref{vuvu}), and obtain
\[\{v_j,b(\m)\}= w_3(u_j-\m)\,b(\m)\,.\]
It together with the equation
\[ 0= \{v_j,b(u_i)\}= \{v_j,b(\m)\}\mid_{\m=u_i}
+b'(u_i)\{v_j,u_i\},\]
gives
\[\{v_j,u_i\} =-{1\over b'(u_i)}\{v_j,b(\m)\}
\mid_{\m=u_i}=\eta\delta_{ij}.\]
Equalities $\{v_i,v_j\}=0$ can be verified in
a similar way:
\ben-\{v_i,v_j\}&=&\{a(u_i),a(u_j)\}
\nn\\
&=&\{a(u_j),a(\m)\}\mid_{\m=u_i}+a'(u_i)\{u_i,a(u_j)\}
\nn\\
&=&a'(u_j)\{a(u_j),u_i)\}+a'(u_i)\{u_i,a(u_j)\}=0.
\nn\en
After the substituting $\l=u_k$ into
the identity
$$a^2(\l)+b(\l)c(\l)=-d(\l)$$
we obtain separation equations
$$v_k^2=-d(u_k)\,,\qquad k=1,\ldots,n\,,$$
where  $d(\l)$ is the determinant of
$L(\l)$.

In the $XYZ$ case we have to use
 entries $\tilde{a}(\l)$ and $\tilde{b}(\l)$
of the similar matrix $\tilde{L}(\l)$
 (\ref{vvop1}).
By the Lemma \ref{lem1} the zeroes of entry
 $\tilde{b}(\l)$
are commuting variables $\{u_i,u_j\}=0$.
For the rest of the proof we refer to \cite{skl86}.

\subsection{Construction of Lax matrices $L(\l)$}
\par\noindent
A general construction of matrices $L_0(\l)$ with
 multiple poles on a direct sum
of loop algebras $\oplus\,{\SL}\left(su(2)\right)$
has been presented
 in detail in \cite{rs87}.
The Lax matrix is assumed in the form
\bq
L_0(\l)=\sum_{\a=1}^3\sum_j^N s^\a_{j}
\,w_\a(\l-\nu_j)\,
\sigma_\a\,,
\label{lres}
\eq
where $\sigma_\a$ are Pauli matrices,
 $w_\a(\l)$ are coefficients of the
corresponding $r$-matrix and $\nu_j$ are  simple poles of
the fixed divisor ${\sl D}=\{\nu_1,\ldots,\nu_N\}$.
 Residues $s^\a_j$ are the standard linear coordinates
on $su(2)^*$ with Lie-Poisson brackets
$$\{s^\a_j,s^\b_k\}=\delta_{jk}
\varepsilon_{\a\b\gamma}\cdot s^\gamma_j$$.
The corresponding phase space is the direct sum of N
 copies $su(2)^*$.
 These matrices describe systems of $N$ interacting
 Euler tops. The case of higher order of poles can be
treated as a degeneration  of (\ref{lres}). It describes
 different tops \cite{rs87} as well.

We shall express matrix elements of $L_0(\l)$ through
 canonically conjugate variables.
Let us consider meromorphic matrix-function $L_0(\l)$
 and the coefficients $w_1(\l-\m)=w_2(\l-\m)$
of the $r$-matrices (\ref{rxxx}-\ref{rxxz}) with one
distinguished point at infinity (it is zero by both
arguments and  their residues at infinity are
equal to $-1$).
 Let us choose
the linear functional $\P_\l$  as the
 higher residue fixed order $K$
at infinity
\bq
\P_\l[z]=-{\rm Res}^K_\infty z(\l)\,.
\label{expphi}
\eq
 Then (\ref{glax} and \ref{vamcom}) gives
\ben
&&-{\rm Res}^K_\infty\left[\dfrac{b(\l)}
{\v(\l-\m)}\right]=u_0\,,
\label{restrentry}\\
&&-{\rm Res}^K_\infty\left[\dfrac{c(\l)}
{\v(\l-\m)}\right]=0\,,
\label{restrentry1}\\
&&-{\rm Res}^K_\infty\left[\dfrac{a(\l)\v'(\l-\m)}
{\v(\l-\m)}\right]=v_0\,,\label{restrentry2}
\en
In particular $u_0=1\,,{~}v_0=0$ for the matrix
 $M_0$ (\ref{vam}).
In the $XXX$ case it means that entries $b(\l)$ and
 $a(\l)$ have the highest in $\l$ terms in power
$K+1$  equal to $u_o$ and $v_0$, respectively. The entry
 $c(\l)$ has  the highest term of order $K$ only.
Note that the conditions  for the entry $a(\l)$
 and $c(\l)$ (\ref{restrentry1}--\ref{restrentry2})
in the Lax pair ($L_0,\,M_0$) (\ref{vam})
follow from the first condition (\ref{restrentry})
for entry $b(\l)$
 and from the definitions $a(\l)=-b_x/2=-\partial_x b/2$
and $c(\l)=-b_{xx}/2$, since the derivative
$\partial_x$
commutes with the functional $\P_\l$.

According to the construction in \cite{rs87} we can fix
 one divisor $\sl D$ consisting of poles  in the
 matrix-function $L_0(\l)$ and a second divisor
 $\sl U$ consisting of zeroes of the entry
$b(\l)$. Let they have the form
\ben
&&{\sl D}=\{(e_j,m_j){~}j=1,\ldots,\tilde{m};\}\,,\nn\\
&&{\sl U}=\{(u_k,n_k){~}k=1,\ldots,\tilde{n};\}\,.
\label{div}\\
&&u_k\neq\infty\,,\quad{\rm and}\quad e_j\neq u_k\,,
{~}\forall j,\,k\,.\nn
\en
where
${\sl D,U}\subset{\bf CP_1}={\bf C}\cup\{\infty\}$
for the $XXX$ model and
${\sl D,U}\subset{\bf CP_1}/2\pi{\bf Z}$ for the
 $XXZ$ model.
We consider the simple zeroes $u_k,{~}n_k\equiv
1;{~}k=1,\ldots,n,$
 only.
This restriction is related to the $r$-matrix structure.
The poles at  $e_j$ can be multiple poles and their
 orders $m_j$
 are fixed.
In the  formulae below each pole is counted according to its
multiplicity and we don't use  special
 indices  when it is clear from a context.

Now we introduce the following ansatz for the entry
$b(\l)$ according  to \cite{rs87,skl86}
\ben
b(\l)&=u_0\cdot \dfrac{\prod_{k=1}^n \v(\l-u_k)}
{\prod_{j=1}^m \v(\l-e_j)}\,,&\nn\\
\label{b}\\
\v(\l)&\equiv w^{-1}(\l)=\l
\qquad&{\rm in{~}the}{~}XXX{~}{\rm case}\,, \nn\\
\v(\l)&=\sinh\l
\qquad&{\rm in{~}the}{~}XXZ{~}{\rm case}\,, \nn
\en
where $u_0$ is either a constant or a dynamical variables.
 The function
$\v(\l)$ depends from the type of the $r$-matrix
and the functional $\P_\l$  (\ref{expphi}) is residue
 with  $K=1$ for $n< m$ and $K=n-m+1$ for $n\geq m$.

The entry $a(\l)$ has $m$ poles at points $e_j$ of the
 divisor $\cal D$ and its  values at $n$ points $u_k$ are
given ($a(\l=u_k)=v_k$) (\ref{sepq}). The asymptotic
behavior at infinity is fixed by Lax representation
(\ref{restrentry}).
 By  Lagrange interpolation formula
the entry $a(\l)$ can be represented as
\ben
&&a(\l)=b(\l)\cdot
\left(v_0 +
\sum_{i=1}^n v_i\cdot w_3(\l-u_i)\cdot g_{ii}\right)\,,\nn\\
\label{a}\\
{\rm where}\quad &&g_{ii}=\left[\left.
\dfrac{\partial b}{\partial\l}
\right|_{\l=u_i} \right]^{-1}=
 {\rm Res}_{u_i}{~} b^{-1}(\l)=
\dfrac{\prod_{j=1}^m \v(u_i-e_j)}{\prod_{k\neq i}^n
\v(u_i-u_k)}
\,,\nn
\en
The $w_3(\l)=\v'(\l)/\v(\l)$ is the coefficient in
the definition of the corresponding $r$-matrix.
Variables $u_0\,,{~}v_0$ are
connected to the Lax matrix $M_0$ (\ref{vamcom}).

Further, from the Lax representation (\ref{vopkdv})
for the initial matrix $L_0(\l)$, we can construct the entry
$a(\l)$. It reads as
\ben
&&a(\l)=-\dfrac{b_x}2=
\dfrac{b(\l)}2\cdot \left(
-\dfrac{\dot{u}_0}{u_0}+\sum_{i=1}^n \dot{u}_i\cdot
w_3(\l-u_i)\right)\,,\nn\\
\label{a1}\\
{\rm where }\qquad &&\dot{u}_k\equiv u_{k,x}=\partial_x u_k\,,
\quad{\rm and}\quad w_3(\l)=\dfrac{\v'(\l)}
{\v(\l)}\,.
\en
We set now, for the simplicity, $u_0=1$
and $v_0=0$ as in the
 Lax matrix (\ref{vam}).
Definition of  entry $c(\l)$ follows
from (\ref{vopkdv})
 as well
\bq
c(\l)=-\dfrac{b_{xx}}2=
-\dfrac{a^2(\l)}{b(\l)}+
\dfrac{b(\l)}2\cdot
\sum_{i=1}^n\left(\ddot{u}_i\,w_3(\l-u_i)
-\dot{u}^2_i \dfrac{\partial w_3(\l-u_i)}
{\partial u_i}\right)\,.
\label{c}
\eq
where $\ddot u_i$ and $\dot u_i$
are defined below.

{}From comparison of two forms of the
entry $a)\l)$ (\ref{a}) and  (\ref{a1})
we can introduce the following Hamiltonian
\bq
H_0=\sum_{i=1}^n v_i^2\cdot g_{ii}=
\sum_{i=1}^n v_i^2\cdot
\left[\left.\dfrac{\partial b(\l)}{\partial
\l}\right|_{\l=u_i}\right]^{-1}\,,
\label{hfree}
\eq
using the Hamilton-Jacobi equations.

The remaining symbols in the entry $c(\l)$ are
defined as
\ben
&&\dot{u}_i\equiv\dfrac{\partial H_0}
{\partial v_i}=2\cdot v_i
 \cdot g_{ii}\,,\nn\\
\nn\\
&&\ddot{u}_i=2\sum_{j=1}^m v_i^2\cdot g^2_{ii}
\cdot\dfrac{\v'(u_i-e_j)}{\v(u_i-e_j)}+\nn\\
\nn\\
&&+2\sum_{k=1}^n \left[(v_k+v_i)^2 g_{ii}g_{kk}-
v_i^2(g_{kk}-g_{ii}) g_{ii}\right]
\v(u_i-u_k)\v'(u_i-u_k)\,,\nn
\en
where $\v(\l)=w_1^{-1}(\l)=\l$ or
 $\v(\l)=\sinh(\l)$.
 Then a direct calculation
shows that the Hamiltonian (\ref{hfree}) is equal to
 the Hamiltonian (\ref{ham})
$H_0=\P_\l[d_0(\l)]$.

The matrix $L_0(\l)$ with entries
(\ref{b}-- \ref{a}) and (\ref{c}) obeys the $r$-matrix
structure
(\ref{rpoi}) as we can check it directly.
 Let us calculate
\ben
\{a(\l),b(\m)\}&=&w_3(\l-\m)b(\m)-w(\l-\m)
b(\l)=\nn\\
\nn\\
-\sum_{k=1}^n\dfrac{g_{kk}\cdot \v'(\m-u_k)}
{\v(\l-u_k)\v(\m-u_k)}\cdot b(\l)b(\m)
&=&
b(\l)b(\m)
\sum_{k=1}^n g_{kk}
\left(\dfrac{w_3(\l-\m)}{\v(\l-u_k)}-
\dfrac{w(\l-\m)}{\v(\m-u_k)}\right)\nn\\
\nn\\
-\sum_{k=1}^n\dfrac{g_{kk}\cdot \v'(\m-u_k)}
{\v(\l-u_k)\v(\m-u_k)}
&=&
\sum_{k=1}^n g_{kk}
\dfrac{\v'(\l-\m)\v(\m-u_k)-
\v(\l-u_k)}{\v(\l-\m)\v(\l-u_k)\v(\m-u_k)}
\nn
\en
The last equlity is true due to the additive
 theorem for $\v(\l)$.

So, we have constructed the initial matrix $L_0(\l)$ in
 the term of variables of separation. {}From the initial
 matrices $L_0(\l)$ we can construct  modified matrices
$L_{MN}(\l)$ (\ref{dopn}). These matrices describe
integrable Hamiltonian system with the Hamiltonian
$H=\P_\l[d(\l)]$ and a complete set of the
 integrals of motion can be
obtained from  determinants of these matrices as well.
The determinant $d(\l)$ is a
meromorphic function with the divisors of poles
 determined
 by divisors of the entries of $L_0(\l)$ or of
$L_{MN}(\l)$.
Therefore the rational algebraic
functions $d(\l)$ can be written  as a quotient of
two product or it can be decomposed into simple fraction
according to the
 Mittag-Leffler theorem
\ben
d(\l)&=&\dfrac{P^N_d(\l)}{Q^m_d(\l)}=
\dfrac{\prod_{k=1}^N\v(\l-h_k)}
{\prod_{j=1}^m \v(\l-e_j)}=\nn\\
&&\label{dhk}\\
&=&P^{N-m}_d\,(\l)+
\sum_{j,i} H_{ji}\cdot\v^{-i}(\l-e_j)=
\prod_{k=1}^{N-m} \v(\l-H_k)+
\sum_{j,i} H_{ji}\cdot\v^{-i}(\l-e_j)\,,\nn
\en
where $N=n$ for the initial matrix $L_0(\l)$. Both
 decomposition provide us with sets of
integrals of motion.
\begin{lemma}
Zeroes $h_k$ of the function $P^N_d(\l)$
can be chosen as a first set of integrals of motion
in involution for the Hamiltonian systems related to the
 Lax matrices $L_0(\l)$ and $L_{MN}(\l)$.
Zeroes $H_k$  of the function $P^{N-m}_d(\l)$
and the residues $H_{ji}$ at  points $e_j$ constitute a
second set of the integrals
of motion in involution.
\label{lem4}
\end{lemma}
{\it Proof:} The proof of involutivity is based on the equality
 (\ref{det}) and
follows the same pattern as the proof of
Theorem \ref{th2}.
Completeness and  functional independence of
 these integrals
have been proved by Mishchenko and Fomenko
(see review \cite{rs87}).

We summarize these considerations as
\begin{prop}
The Lax matrix $L_0(\l)$
(\ref{b}, \ref{a}, \ref{c}) describes
geodesic motions  on the
Riemannian manifold with the
complex diagonal meromorphic Riemannian metrics
\bq
g^{kj}
=\delta^j_k\cdot
\left[\left.\dfrac{\partial b(\l)}{\partial \l}
\right|_{\l=u_i}\right]^{-1}\,.
=\delta^j_k\cdot\dfrac{\prod_{j=1}^m \v(u_i-e_j)}
{\prod_{k\neq i}^n \v(u_i-u_k)}\,.
\label{metric}
\eq
defined on  moduli of  $n$-dimensional Jacobi varieties
in terms of complex elliptic (root) coordinates $u_k$.
The Hamiltonian system (\ref{hfree})
$$H_0=\sum_i^n v_i^2 g_{ii}$$
has a complete system of
first integrals $v_k$ and it defines a Lagrangian
submanifold of the phase space ${\bf C}^{2n}$
which has the form of  symmetric product
$((\Gamma\times\ldots\times\Gamma)/\sigma_n)$
of  $n$ copies of Riemannian surfaces \cite{almar92}
\bq
\Gamma:\quad v_k^2=-\left.d(\l)\right|_{\l=u_k}
=-\left[\,\dfrac{P_d(\l)}{Q_d(\l)}\,\right]_{\l=u_k}
\eq
A modified Lax matrix $L_{MN}(\l)$ (\ref{dopn})
describes
 potential motion on the same Riemannian manifold
with a Hamiltonian of
natural type
\bq
H=H_0+V_{MN}\,(u_1,u_2\ldots,u_n;f_{-M},f_{-M+1},\ldots,f_N)\,,
\eq
 $f_j$ are coefficients of the function $f(\l)=
\sum f_k\l^k$ or of $f(\l)=\sum f_k e^{k\l}$.
These matrices obey the $rs$-matrix structure (\ref{rspoi})
with the $s_{MN}$-matrix given
by the Theorem \ref{theorem2}.
\end{prop}
In the potential case we reconstruct, for instance,
 motions under permutationals symmetric
potential classically
(via point transformation) separable in
generalized elliptic coordinates
\cite{wo85}. For $m\leq n+1$ we have here
the  well known Jacobi
 problem of geodesic motions on quadrics
\cite{al81,arw92}, corresponding quadrics
are defined by condition
 (\ref{restrentry})
$$-{\rm Res}_\infty^K\left[\dfrac{b(\l)}{\v(\l-\m)}
\right]=1\,,
\quad\Longleftrightarrow\quad
 Q^n\left\{\sum \dfrac{x_j^2}{e_j}=1\right\}$$
with  variables $x_j$ will be introduced in section
4.1.

For these systems we can define a Riemann matrix
 of periods of
 holomorphic differentials
and by using the Abel-Jacobi map we can find
the action-angle
 variables. The potentials $V_{MN}$ are the
finite gap potentials
on these manifolds \cite{dnm76,al81}.
For details about
geodesic and potential motion with metric
(\ref{metric}) in the $XXX$ case
see \cite{al81,al92,almar92}.

Thus we have constructed  initial matrices
$L_0(\l)$ and modified matrices $L_{MN}(\l)$
(\ref{dop}--\ref{dopn}) which
correspond to a  geodesic motion  and a potential
 motion on
Riemannian manifolds with metric (\ref{metric}),
respectively.
For the $XXX$ case the initial matrix $L_0(\l)$
(\ref{b}, \ref{a}, \ref{c})
was considered in \cite{kkm94} and the
Hamiltonian $H_0$ (\ref{hfree})
describes free motion on Riemannian spaces of
constant curvature.

The above results describe motions on Jacobi variety
$((\Gamma\times\ldots\times\Gamma)/\sigma_n)$
which is a symmetric product of  uniform Riemannian surfaces
\bq
\Gamma:{~}v_k^2=-\left.d(\l)\right|_{\l=u_k}\,.
\eq
This scheme can be generalized to the case of the
  more general
Jacobi variety
$((\Gamma_1\times\Gamma_2\ldots\times\Gamma_n))$
with  different Riemannian surfaces
\bq
\Gamma^{(j)}_k:{~}v_k^{(j)\,2}=-\left.d^{(j)}(\l)
\right|_{\l=u_k}\,,\qquad
k=1,\ldots,n_j,{~}j=1,\ldots,n\,.
\eq
where $k$ and $j$ are two numbers indexing curves.
Motion on such manifolds can be described with the use of
the block matrices
\bq
L(\l)=\oplus^{n}_j\,
L^{(j)}(\l,v^{(j)}_1,u^{(j)}_1;\ldots;v^{(j)}_{n_j},
u^{(j)}_{n_j})\,,
\label{lkj}
\eq
acting in the auxiliary space
\bq
V_{aux}=\oplus_{j} V^{(j)}_{aux}\,.
\eq
Here $j$ numbers  blocks and $k$  denotes number of degree
 of freedom related to
each block.
Each block $$
L^{(j)} (\l, v^{(j)}_1, u^{(j)}_1; \ldots; v^{(j)}_{n_j},
 u^{(j)}_{n_j})$$
is a matrix of  type $L_0(\l)$ or
$L_{MN}(\l)$ (\ref{dop}), respectively.
These manifolds have been investigated in \cite{van92}
and such
 block matrices
were considered in \cite{rav93,ts94}.

In the $XYZ$ case  parameterization of the entry $b(\l)$
is  more complicated. For instance, it has to comply with
the equality (\ref{zerob})
$$\{\tilde{b}(\l),\tilde{b}(\m)\}=
g(\l,\m)\cdot\tilde{b}(\l)-g(\m,\l)\cdot\tilde{b}(\m)\,.$$
In order to bypass this difficulty we can use two-dimensional
lattice averaging \cite{ft87} for the constructing
matrices $L_o(\l)$
 from the corresponding matrices of the rational model.

In  quantum mechanics the metric (\ref{metric})
defines standard
operator of the Laplace-Beltrami type and stationary
 Schr\'odinger
equations
$$\nabla^j\nabla_j \Psi+
\left(\,E-V_{MN}\,(u_1,u_2\ldots,u_n;f_{-M},\ldots,f_N)\,
\right)\Psi=0$$
on the $n$-dimensional Riemannian manifolds
 \cite{al92,almar92}.
 Quantum separation equation for  geodesic motion and
potential motion  and their properties have
 been considered  in
\cite{al92,almar92} but not in the $r$-matrix approach.
In the framework of the linear $r$-matrix formalism
separation of variables has a direct quantum
counterpart which
is described
in \cite{skl86,skl89,skl92}. In the quantum case we replace
variables $u_i,\,v_i$ by operators and
Poisson brackets (\ref{rpoi}) by the commutators
$$\left[\on{L}{1}(\l),\on{L}{2}(\m)\right]=
[r(\l,\m),\on{L}{1}(\l)+\on{L}{2}(\m)\,]\,.$$
Note that dynamical $s_{MN}(\l,\m)$ matrices depend on
 coordinates only.
Separated equations have the following form
$$\left(\dfrac{d^2}{d\,u_k^2}+
U(u_k;\,e_1,\ldots,e_m;\,f_{-M},\ldots,f_N;h_1,
\ldots,h_n)\right)
\psi_k(u_k,h_1,\ldots,h_n)=0$$
where potential
$U(u_j)$ is a rational algebraic function of the
 variables $u_j$ parametrized by
 $e_i,{~}i=1\ldots m;$ and $f_j,{~}j=-M\ldots N$ and
$h_j$ are the eigenvalues integrals of motion.
According to \cite{skl89}, common multiparticle
eigenfunction of  quantum integrals of motion is equal to
$$ \Psi(u_1,\ldots,u_n)=\prod_{k=1}^n\, \psi_k\,(u_k)\,.$$
A quantum  operator $L_0(\l)$
for  free motion in the $XXX$ case was investigated in
\cite{skl89,kuz92a,kkm94} while,
the quantum modified operator $L(\l)$ for potential
 motion
was considered in \cite{eekt94,krw94}.

\section{Canonical transformations of variables}
\setcounter{equation}{0}
We can represent  meromorphic matrix-functions $L(\l)$
by means
of  generators of various Lie groups, which one
can  consider as the various phase spaces for the
integrable systems.
This leads to links among  different integrable
systems: the Neumann problem, Gaudin magnets
 and Euler tops
 \cite{rs87,kkm94}. In this paper we shall consider
finite-dimensional integrable systems of  natural type
and shall investigate
three types of  canonical transformations of variables
prescribed by  linear $r$-matrix structure
(\ref{rpoi},\ref{rspoi}).

\subsection{Generalized elliptic coordinates}
If the limit orders of poles $m_j$ at the points $e_j$ are
fixed then the
Poisson brackets at  different poles are independent and
the corresponding
orbits are a direct product of  orbits of algebra $su(2)$.
This factorization
is independent of the points $e_j$ \cite{rs87}.
By the Mittag-Leffler theorem we can decompose
rational functions
$b(\l)$ (\ref{b}) and $a(\l)$ (\ref{a})
into partial fractions
\ben
b(\l)&=&P^{n-m}_b\,(\l)+\sum_{j,i} b_{ji}
\cdot\v^{-i}(\l-e_j)=
\prod_{k=1}^{n-m} \v(b_k-\l)+\sum_{j,i} b_{ji}\cdot\v^{-i}
(\l-e_j)\,,\nn\\
\label{fab}\\
a(\l)&=&P^{n-m-1}_a\,(\l)+\sum_{j,i} a_{ji}\cdot\v^{-i}(\l-e_j)=
\prod_{k=1}^{n-m-1} \v(a_k-\l)+\sum_{ji} a_{ji}\cdot\v^{-i}(\l-e_j)
\,,\nn
\en
where $b_k$ and $a_k$ are  zeroes of the polynomials
$P^{n-m}_b(\l)$
and $P^{n-m-1}_a(\l)$, respectively.
The residues $b_{ji}$ and $a_{ji}$ at the  points $e_j$ are
linear coordinates on the algebra $su(2)^*$ (\ref{lres}).
We use later  two representations for
generators of this algebra
in terms of canonical variables
\ben
b_j=x_j^2\,,\qquad &a_j=x_jp_j\,,\qquad &c_j=p_j^2\,,
\label{ell}\\
b_j=2x_jX_j\,,\qquad &a_j=x_jp_j+X_jP_j\,,
\qquad &c_j=2p_jP_j\,.\label{ell2}
\en
We omit here the double indexes $ji$ which are
 necessary for higher order
 poles and we used  notation $c_j$ for
residues of entry $c(\l)$.
Variables $x_j\,,{~}p_j$ and $X_j\,,P_j$
are canonically
 conjugate coordinates and momenta.
Generalized elliptic coordinates correspond
to the first
 representation
(\ref{ell}) and  in our scheme they are a simple consequence
 of the Mittag-Leffler
 theorem for meromorphic functions.

The other $(n-m)$ variables can be chosen as  zeroes of the
polynomial $P_b^{n-m}(\l=x_k)=0$ and as the values of function
$P_a^{n-m-1}(\l)$ at these zeroes $x_k$
\bq
P_b^{n-m}(\l=x_k)=0\qquad p_k=P_a^{n-m-1}(\l=x_k)
\label{dell}
\eq
Variables $x_k,\,k=m+1,\ldots,n$ are calculated from the
 coefficients
of the polynomial $P^{n-m-}_b(\l)=\sum B_k\l^k$.
For instance in the $XXX$ case with simple poles these
 coefficients are equal to
\bq
B_k=\left.\dfrac1{(k-1)!}\,\dfrac{d^{k-1}}{d\l^{k-1}}\,b(\l)
\right|_{\l=0}+(-1)^k\sum_{j=1}^m\dfrac{x_j^2}{e_j^{k+1}}\,,
\eq
then the conjugate (to the $x_k$) momenta $p_k,{~}k=m+1,
\ldots,n$ are
\bq
p_k=\left.a(\l)\right|_{\l=x_k}-\sum_{j=1}^m
\dfrac{x_j\,p_j}{x_k-e_j}\,.
\eq
\begin{lemma}
Variables $x_j,p_j{~}j=1,\ldots,m$ and $x_k,p_k{~}k=m+1,\ldots,n$
are a system of canonically conjugate variables.
\label{lem5}
\end{lemma}
{\it Proof} The Mittag-Leffler theorem defines a unique,
invertible map between old
$n$ pairs of variables $(u_k,v_k),{~}k=1,\ldots,n$ and the
new $n$ pairs of
variables $(x_k,p_k),{~}k=1,\ldots,n$.
The proof of canonical conjugation  of these variables
is based on
the $r$-matrix structure. The technique developed
by Sklyanin (see \cite{skl86} and proof of the
Theorem \ref{th2})
 has to be employed.

In the case of higher order poles we have an additional freedom
in representing  residues through canonical variables. There
 are two approaches to this problem.
The first one is connected with  standard degenerations of
simple divisors, for example
$e_j\to e_{j+1}$ or $e_j\to \infty$. This approach
is developed
 in \cite{kkm94,kuz92a}.
Our approach is different.

We shall show how the variables $(q_j,p_j)$ are connected to
Newton type variables for stationary flows of KdV
 type equations
 introduced in \cite{rw92}. We shall consider the $XXX$ case
 but generalization
for the $XXZ$ case is transparent. The $r$-matrix algebra is
 associative;
a linear combination of any two matrices $L_1$ and $L_2$ satisfying
 (\ref{rpoi}) satisfies (\ref{rpoi}) too.
Let us consider the
polynomial part of $L_0(\l)$ corresponding to poles at infinity.
The determinants of $L_0(\l)$ and $L(\l)$ from
 (\ref{veqmot}--\ref{deqmot}) are
\ben
d_0(\l)&=&\dfrac{b\cdot b_{xx}}{2}-
\dfrac{b_x^2}{4}\,,\label{dob}\\
\nn\\
d(\l)&=&\dfrac{b\cdot b_{xx}}{2}-
\dfrac{b_x^2}{4}+b^2\cdot u_{MN}(\l)\,.\label{ddb}
\en
They are integrated forms of the KdV recursion relations
 \cite{rw92}.

A simple substitution for the entries of matrix $L_0(\l)$
\ben
b(\l)&=&{\cal B}^2\,,\qquad
a(\l)=-b_x/2=-{\cal B}{\cal B}_x\,,\label{subsb}\\
c(\l)&=&-b_{xx}/2=-{\cal B}_x^2-{\cal B}{\cal B}_{xx}\,,\nn
\en
turns determinants (\ref{dob}--\ref{ddb}) into the form
\bq
d_0(\l)={\cal B}^3{\cal B}_{xx}\,,\qquad
d(\l)={\cal B}^3{\cal B}_{xx}+{\cal B}^4
\left[\dfrac{f(\l)}{{\cal B}^4}\right]_{MN}\,,\label{dsub}
\eq
if we use an explicit formula for the potential $u_{MN}(\l)$.
These equations  have the form of Newton equations for $\cal B$
\ben
{\cal B}_{xx}=d_0(\l){\cal B}^{-3}\,,\nn\\
\label{newton}\\
{\cal B}_{xx}=d(\l){\cal B}^{-3}-{\cal B}
\left[\dfrac{f(\l)}{{\cal B}^4}\right]_{MN}\,,\nn
\en
If we assume that ${\cal B}=\sum_{j=0}^N q_{N-j}\,
\l^j$ is a polynomial then its coefficients $q_j$
 obey the Newton equation of motion
(\ref{newton}) with $d(\l)=\sum I_k\l^k$,
 where $I_k$ are integrals of motion.
It becomes clear in terms of variables $q_j$ that deformation
 of integrals of motion $I_k$ affects only the potential
($q$-dependent) part. The kinetic (momentum dependent)
 part of $I_k$ remains unchanged.

 Coefficients of all entries $a(\l)\,,{~}b(\l)$
and $c(\l)$ are easily expressed in terms of coefficients
 of $\cal B$. Let us fix the highest order of $\l$
in entry as $K$, then
\ben
&&b(\l)=\sum_{j=0}^K b_j\l^j\,,\qquad
a(\l)=\sum_{j=0}^K a_j\l^j\,,\qquad
c(\l)=\sum_{j=0}^K c_j\l^j\,,\label{sets}\\
\nn\\
{\rm with}\qquad&& b_K=1\,,\quad a_K=0\,, \quad c_K=0\label{br}
\en due to the definition of $M_0$ and $\P_\l$. Let
$${\cal B}(\l)=\sum_{j=0}^K q_{K-j}\l^j\,,{~}{~}
{\rm with}{~}{~}q_0=1\,,$$
where the set $\cal B$ we consider as a formal set.
After substitution of these definitions into
 (\ref{subsb})
we obtain
\ben
b_j&=&\sum_{i=0}^{K-j} q_i\, q_{K-j-i}\,,\qquad
a_j=-\sum_{i=0}^{K-j} q_{i,x}\, q_{K-j-i}\,,\nn\\
\label{ent}\\
c_j&=&-\sum_{i=0}^{K-j} q_{i,x}\,q_{K-j-i,x}
-\sum_{i=0}^{K-j} q_{i,xx}\,q_{K-j-i}\,,\nn
\en
where $q_x$ is  derivate with respect to  $x$ and we
 used the Newton formulae for a product of sets.
{}From $q_0=1$ and the definitions (\ref{ent}) the
 restrictions (\ref{br}) for the higher coefficients
 of  entries of matrix $L_0$ follow automatically.
Canonically conjugate (to the coordinates $q_j$)
momenta $p_j$ can be derived from the $r$-matrix algebra
 (\ref{rpoi}) and from the definitions (\ref{sets}--\ref{ent})
$$p_j=q_{K+1-j,x}\,,\qquad\{p_j,q_k\}=\delta_{jk}\,.$$
 As an example, we present first polynomials
\ben
K=1\,,\quad b(\l)&=&\l+2q_1\,,\qquad -a(\l)=p_1\,,\nn\\
K=2\,,\quad b(\l)&=&\l^2+2\l x_1+(2x_2 +x_1^2)\,,\nn\\
-a(\l)&=&\l p_2+(p_1 +p_2x_1)\,,,\label{nrep}\\
K=3\,,\quad b(\l)&=&\l^3+2\l^2 x_1+\l (2x_2 +x_1^2)
+2(x_3+x_1x_2)\,,\nn\\
-a(\l)&=&\l^2 p_3+\l (p_2 +p_3x_1)
+(p_1+p_2x_1+p_3x_2)\,,\nn
\en
All integrals of motion can be expressed in  terms of new
variables directly from the equations (\ref{dsub}) by taking
 residues. Notice that the kinetic part of the Hamiltonian
 $H=\P_\l[d(\l)]$ has the nondiagonal form
\bq
T=\sum_{j=1}^{K+1}p_jp_{K-j}\,.\label{kin}
\eq

Definition of variables of separation is not changed by
 deforming the Lax matrix
 $L_0(\l)$ (\ref{dop}). For instance, the free motion
 Hamiltonian $H=T$ (\ref{kin}) and the corresponding Hamiltonian
 with potential terms $H=T+V_{MN}$ separate in
the same coordinate
systems.
Later we shall show some examples illustrating the above scheme.

\subsection{ Jacobi coordinates}
Let us take a simplest  matrix $L^{(k)}_0(\l)$
with the entry $b(\l)$ of (\ref{b}) which has only
 one zero
\bq
L^{(k)}_0\,(\l,v_k,u_k)=
\left(\begin{array}{cc}-v_k&\v(\l-u_k)\\0&v_k\end{array}\right)\,.
\label{lvmat}
\eq
It obeys the linear $r$-matrix algebra (\ref{rpoi})
with the $r$-matrix (\ref{rxxx}) or  (\ref{rxxz}).
Let
\bq
L_0(\l)=\oplus^n_k\, L_0^{(k)}(\l,v_k,u_k)
\eq
what means that we associate to our system  a matrix $L$, which
 has  blocks structure
and each block $L^{(k)}_0$ obeys the $r$-algebra with the same
$r$-matrix.

A modified matrix $L(\l)$ (\ref{dop}) is equal to
\bq
L(\l)=\oplus^n_k\, L^{(k)}_{MN}(\l,v_k,u_k)\,,
\label{lpr}
\eq
where  blocks  $L^{(k)}_{MN}(\l,v_k,u_k)$ are
\ben
L^{(k)}_{MN}(\l,v_k,u_k)&=&
\left(\begin{array}{cc}-v_k&\v(\l-u_k)\\
\left[\dfrac{f(\l)}{\v(\l-u_k)}\right]_{MN}&v_k\end{array}\right)=
\nn\\
\label{lnk}\\
&=&\left(\begin{array}{cc}
-v_k& \v(\l-u_k)\\c_{MN}(\l,u_k)&v_k\end{array}\right)\,.\nn
\en
These matrices are  deformations of the initial matrices
$L_0(\l,v_k,u_k)$ (\ref{lvmat}) and
each block $L^{(k)}_{MN}$ obeys the $rs$-algebra with a common
$r$-matrix but different (since they depend on $u_k$) matrices
 $s^{(k)}_{MN}$ constructed according to (\ref{alpn}).

The matrix  $L_0$ (\ref{lvmat}) and the modified
 matrix $L_{MN}$ (\ref{lnk})
have a  internal structure \cite{kuz92a}
\ben
L_0\,(\l,v_k,u_k)&=&
\left(\begin{array}{cc}-v_k&\v(\l-u_k)\\0&v_k\end{array}\right)=
\left(\begin{array}{cc}-\sum \a_j p^{(k)}_j&\v(\l-\sum \a^{-1}_j
 q^{(k)}_j)\\
0&\sum \a_j p^{(k)}_j\end{array}\right)\nn\\
&&\label{str}\\
L_{MN}(\l,v_k,u_k)&=&
\left(\begin{array}{cc}-\sum \a_j p^{(k)}_j&\v(\l-\sum\a^{-1}_j
q^{(k)}_j)\\
\left[\dfrac{f(\l)}{\v(\l-\sum\a^{-1}_j q^{(k)}_j)}\right]_{MN}&
\sum \a_j p^{(k)}_j\end{array}\right)\nn
\en
if variables $v_k$ and $u_k$ are considered as
 linear combinations
of certain canonical variables
$p^{(k)}_j,\,q^{(k)}_j{~}j=1,\ldots K$.
If we consider a system with of  $n$ degree of
freedom  and require that
 Hamiltonian  has a natural canonical form with
the kinetic
energy form
$$T^{(n)}=\sum_{k=1}^n v_k^2=\sum q_j^2$$
then  the internal structure of
(\ref{str}) leads to the Jacobi transformations for
an $n$-degrees of freedom systems. It is then a pure
 coordinate change of
variables.
As before free motion equations  corresponding to the
 undeformed matrix and
motion with potential corresponding to  modified matrix
 separate in same system of coordinate.

\subsection{Momentum dependent change of variables}
Let us consider
functions $B_j(\l)$ and $A_j(\l)$,
which are  algebraic functions
of  entries of the matrix $L(\l)$ and
which satisfy   suitable Poisson brackets, for example
\ben
&&\{A_j(\l),A_k(\m)\}=\{B_j(\l),B_k(\m)\}=0\,,\nn\\
&&\{A_j(\l),B_k(\m)\}=\delta_{jk}\cdot g(\l,\m)
\cdot\left(B_j(\l)-B_k(\m)\right)\,.\nn
\en
We shall require that zeroes of the functions $B_j(\l)$
and values of  functions $A_j(\l)$ at the these zeroes
define a new  system of coordinates.
Because  functions $B_j(\l)$ and $A_j(\l)$ are
meromorphic functions it will also be  interesting to
 consider the corresponding
residues. As an example, we present such transformation
 related to Henon-Heiles system \cite{ts94}.

Let us restrict to the $XXX$ case.
In same way as for the Jacobi transformations
we consider the  block matrices $L(\l)$ (\ref{lpr}) with
the simplest blocks (\ref{lnk}) \cite{ts94}.

\begin{lemma} The map $(u,v)\longrightarrow (x,p)$ given by
\ben
x_k^2&=& \a\left.\dfrac{d_k(\l)-d_{k+1}(\l)}
{b_k(\l)-b_{k+1}(\l)}\right|_{\l=0}\,;
\quad \a\in {\bf R}\,,\quad k=1,\ldots,n-1\,,\nn\\
\nn\\
p_k&=&
\left.\dfrac{a_k(\l)-a_{k+1}(\l)}{b_{k}(\l)
-b_{k+1}(\l)}\right|_{\l=0}
\cdot x_k\,;\nn\\
&&\label{zam}\\
x_n&=&\b\left.\sum_{k=1}^n b_k(\l)\right|_{\l=0}-
\sum_{k=1}^{n-1} \dfrac{\a\b}2\left(\dfrac{p_k}{x_k}\right)^2\,,
\qquad \b\in {\bf R}\,,\nn\\
\nn\\
p_n&=&\dfrac{1}{\b}\left.\sum_{k=1}^n a_k(\l)\right|_{\l=0}
+\a\b\sum_{k=1}^{n-1}
\dfrac{p_k}{x_k}\cdot\left(1+\dfrac{p_k^2}{x_k^2}\right)\,;\nn
\en
where $a_k(\l)$ and $b_k(\l)$ are the entries of  matrices
$L_{MN}(\l,v_k,u_k);{~}k=1,\ldots,n$ (\ref{lnk}) and $d_k(\l)$
 are the corresponding determinants,
is a canonical transformation.
\label{lem6}
\end{lemma}
{\it Proof.} We fix the variable $x_k^2$ and the Hamiltonian
$H=\sum_{k=1}^n d_k(\l=0)$.
Then a corresponding momentum $p_k=\{H,x_k\}$ follows from
the $rs$-algebra (\ref{rspoi})
\ben
2 x_k p_k&=&\{H,x_k^2\}\nn\\
&=&
\a\left.
\left\{d_k(\l)+d_{k+1}(\l),
\dfrac{d_k(\m)-d_{k+1}(\m)}{b_{k}(\m)-b_{k+1}(\m)}\right\}
\right|_{\m=0,\l=0}\nn\\
&=&
-\a \dfrac{d_k(\m)-d_{k+1}(\m)}{(b_{k}(\m)-b_{k+1}(\m))^2}\cdot
\left.
\left\{d_k(\l)+d_{k+1}(\l),b_{k}(\m)-b_{k+1}(\m)\right\}
\right|_{\m=0,\l=0}\nn\\
&=&
2\a \left.
\dfrac{d_k(\m)-d_{k+1}(\m)}{b_k(\m)-b_{k+1}(\m)}\cdot
\dfrac{a_k(\m)-a_{k+1}(\m)}{b_k(\m)-b_{k+1}(\m)}
\right|_{\m=0}\nn\\
&=&2\left.\dfrac{a_k(\m)-a_{k+1}(\m)}{b_k(\m)-b_{k+1}(\m)}
\right|_{\m=0}\cdot x^2_k\nn
\en
where we have used the explicit form of the $L_{MN}$
matrices (\ref{lnk})
and the technique developed by Sklyanin \cite{skl89}.
The variable $x_n$ is fixed by the condition $\{x_k,x_n\}=0$
 and
 the corresponding
momentum $p_n=\{H,x_n\}$ is calculated from the $rs$-algebra
and from
the Hamilton-Jacobi equations.

By the proof we have used  the following relations
\ben
\{b_k(\l),a_j(\m)\}&=&\dfrac{\delta_{kj}}{\m-\l}
(b_k(\m)-b_k(\l))\,,\nn\\
\{b_k(\l),c_j(\m)\}&=&\dfrac{2\delta_{kj}}{\m-\l}
(a_k(\m)-a_k(\l))=0\,,\nn\\
\{b_k(\l),b_j(\m)\}&=&0\,,\nn
\en
which are determined by the $r$-matrix only and
the one relation
$\{d_k(\l),d_j(\m)\}=0$, which is given the $rs$-algebra
(\ref{rspoi}) with $s_{MN}^{(k)}$-matrices (\ref{alpn}).

We emphasize that it is a non-pure coordinate change of variables
and that a free motion on manifold and a potential motion on it
 separate in  different systems of coordinates.
 The canonical
 transformation (\ref{zam}) is a technical result as yet, but it
 is known that the nonlinear equations behind
 the particular systems
separated in these variables and in the Jacobi variables are
 related by a Miura map \cite{fo91}.

\subsection{Examples}
We restrict ourselves to the $r$-matrix algebra of the $XXX$ type
and shall consider in detail only the  Taylor projection
 (\ref{cut+}).

 As an example, we discuss here the
Henon-Heiles system and a quartic system of two degrees of freedom.
 Their potentials
are  polynomials of degree three and four, respectively.
The Henon-Heiles potential reads
($A,\,B,\,C$ are constant):
$$
H=\dfrac12 (p_x^2+p_y^2)+\dfrac12(Ax^2+By^2)+x^2y+C y^3\,.
$$
It has been extensively studied both in nonintegrable and
 integrable regimes.
It is known \cite{fo91,pe91} to be integrable for the following
 three sets of parameters
\ben
(i)\qquad &A=B,\qquad &C=1/3\nn\\
(ii)\qquad &A,B{~}{~}{\rm arbitrary} &C=2\label{hh}\\
(iii)\qquad &16A=B,\qquad &C=16/3\nn
\en
The quartic potential is described by
$$
H=\dfrac12 (p_x^2+p_y^2)+Ax^4+Bx^2y^2 +C y^4\,.
$$
It is known \cite{pe91} to be integrable for the following three
sets of parameters
\ben
(i)\qquad &B=6A,\qquad &C=A\nn\\
(ii)\qquad &B=12A,\qquad &C=16A\label{quart}\\
(iii)\qquad &B=6A,\qquad &C=8A\nn
\en
Separation of variables of these Hamiltonians takes place in the
coordinate systems
defined in sections 4.1--4.3.

The type ($i$) systems are related to the Jacobi coordinates
and  to the block matrices $L(\l)$ (\ref{lvmat}, \ref{lpr}).
Let us consider entries $c_N(\l,u_k)$ in matrices
$L_N(\l,u_k,v_k)$ (\ref{lnk}) which are the following polynomials
 of two variables
$\l$ and $u_k$:
\bq
c_N(\l,u_k)=
\left[\sum^N_{i=0}f^{(k)}_{i}\l^{i}
\cdot\sum^{+\infty}_{j=0}\dfrac{u_{k}^j}{\l^{j+1}}\right]_+
=\sum^{N-1}_{i=0}\l^{i}\,\sum^{N-1-i}_{j=0} u_k^j f^{(k)}_{j+i+1}\,.
\label{cn}
\eq
We remind that brackets  $[\,\cdot\,]_+$ denote a Tailor projection
(\ref{cut+}).
Determinants $d^{(k)}_N$ of  matrices
$L^{(k)}_N(\l,u_k,v_k)$ (\ref{lnk}) are equal to
\bq
d^{(k)}_N=\sum^{N}_{i=1}f^{(k)}_{j}\,(u_k^j-\l^j)-v_k^2\,.
\label{deln}
\eq
A generating function of  integrals of motion can be taken
as a determinant of the matrix (\ref{lpr})
$d(\l)\equiv\det L(\l)=\prod _{k=1}^n d^{(k)}_N$.
Hamiltonians for these systems can be defined by
\bq
H^{n}_{N}=\left.\dfrac{d^{(n-1)N}}{d \l^{(n-1)N}}\,d(\l)
\right|_{\l=0}\,,
\label{hamilt}
\eq
and their explicit form
up to a constant factor reads
\ben
H^{n}_{N}&=&
\sum_{k=1}^{n}\left[ f_N^{(k)}\right]^{-1}\,v_k^2 \,
-\,\sum_{k=1}^{n}\left[ f_N^{(k)}\right]^{-1}\,
\sum_{j=1}^N f_j^{(k)}
\cdot u_k^j\nn\\
&=& T^{(n)}+\b V^{(n)}_N\,, \qquad \b\in {\bf R}\,,
\label{ham+}
\en
where $T^{(n)}$ denotes kinetic energy and $V^{(n)}_N$
 is a potential.
If higher coefficients $f^{(k)}_N$ of the polynomials
 $f^{(k)}(\l)$
are  the same for all degrees of freedom
$f^{(k)}_N=f^{(j)}_N{~}{\rm for}{~}\forall k,j\,$
then the hamiltonians
(\ref{hamilt}) can be rewritten  as
\bq
H^{n}_{N}=\left.\sum_{k=1}^{n}d_k(\l)\right|_{\l=0}\,.
\label{hamd+}
\eq
In the case of the Laurent projection $[\,\cdot\,]_{MN}$
 (\ref{cutmn})
Hamiltonians (\ref{ham+}) are equal to
\bq
H^{n}_{N}=
\sum_{k=1}^{n}\left[ f_N^{(k)}\right]^{-1}\,(v_k^2-
\sum_{j=-(M-1)}^N f_j^{(k)}\cdot u_k^j\,)\,,
\eq

Now by exploiting the internal structure of the matrix $L(\l)$
and by using the respective Jacobi transformations (\ref{str})
we shall illustrate the general scheme through the  cases of
two and three degrees of freedom.
For two degrees of freedom after the transformation
$u_1=q_1+q_2$, $u_2=q_1-q_2$ the homogeneous potentials
$V^{(2)}_N$ of
degrees $j=1,2,\ldots N$ read:
\ben
V^{(2)}_1&=&e^+_1\,q_1+e^-_2\, q_2\,;\nn\\
V^{(2)}_2&=& V^{(2)}_1+ e^+_2\,(q_1^2+q_2^2)+e^-_2\,q_1q_2\,;\nn\\
V^{(2)}_3&=&V^{(2)}_2 + e^+_3\,(q_1^3+3q_1q_2^2)+e^-_3\,(q_2^3+
3q_1^2q_2)\,;\nn\\
&&\label{pot2}\\
V^{(2)}_N&=&\sum_{j=1}^{N}\,\sum_{i=0}^{j}C_j^i
(f_j^{(1)}+(-1)^i\cdot f_j^{(2)}\,)q_1^{j-i}\cdot q_2^i\,;\nn
\en
where $e^{\pm}_N=(f^{(1)}_N\,\pm f^{(2)}_N\,)$ and
$C^i_j$ are the binomial coefficients.

For  three degrees of freedom and $T=\sum p_k^2$ we can choose
the coefficients $f_N^{(k)}$ in such way that the Jacobi
transformation has the simplest possible form
$$u_1=(q_1-2q_2+q_3)\,,\qquad u_2=-3(q_1-q_3)\,,
\qquad u_3=(q_1+q_2+q_3)$$
with $f^{(1)}_N=6$, $f^{(2)}_N=18$, $f^{(3)}_N=3$.
The first homogeneous potentials $V^{(3)}_N$ of
degree $j=1,2,\ldots N$
are
\ben
V^{(3)}_1&=&(f^{(1)}_1-3f^{(2)}_1+f^{(3)}_1)q_1
+ (f^{(3)}_1- 2f^{(1)}_1)q_2 + (f^{(1)}_1+3f^{(2)}_1
+f^{(3)}_1)q_3\,;\nn\\
V^{(3)}_2&=& V^{(2)}_1+(f^{(1)}_2 + 9f^{(2)}_2 +
 f^{(3)}_2)(q_1^2 +q_3^2)
+ (4f^{(1)}_2+f^{(3)}_2)q_2^2 \nn\\
&+& (2f^{(3)}_2-4f^{(1)}_2)(q_1q_2+q_2q_3)
+ 2(f^{(1)}_2- 9f^{(2)}_2 +f^{(3)}_2)q_1q_3\,;\nn\\
&&\label{pot3}\\
V^{(3)}_N&=&V^{(3)}_{N-1}+{~}
\sum_{j+k+l=N}{~}f_{jkl}\,q_1^{j}\,q_2^k\,q_3^l\,;\nn
\en
where the $f_{jkl}$ coefficients are expressed
through $f_j^{(k)}$  and the binomial coefficients $C^i_j$.
For two degrees of freedom the
Hamiltonians $H_3^{(2)}$ and $H_4^{(2)}$ (\ref{ham+},\ref{pot2})
coincides with the Hamiltonians of the
Henon-Heiles system and of the quartic system of type ($i$).

Systems of type ($iii$) are described by the
same block matrices $L(\l)$ and by the same function
$f_{MN}(\l)$ as type ($i$) systems. Systems of type ($i$)
 are separated in
Jacobi systems of coordinates while systems of type
 ($iii$ separate
after non-pure coordinate canonical transformation of variables
described in Lemma \ref{lem6}.

The type ($ii$) systems are separated
in generalized elliptic coordinates. The corresponding
matrices $L(\l)$ were introduced in \cite{krw94,eekt94}
and have been investigated in detail there.

 Hamiltonians (\ref{hh}) and (\ref{quart})
correspond to a special choice of the function $f_N(\l)$;
for the Henon-Heiles system
\bq
f_3(\l)=
-\dfrac16(\l^3-\dfrac12 A\l^2-\dfrac32 A^2\l)\,.
\eq
The corresponding particular Lax equations and separation of variables
were studied in \cite{rav93} for the special choice of the
function $f_N(\l)$,  variables of separation for particular
 quartics potentials were discussed in \cite{rav94}.
Some extensions of these system are known \cite{pe91}. They
contain  more general functions $f_{MN}(\l)$ and
can be investigated within our scheme which describes the
 infinite family of rational potentials.

\section{Conclusions}
\setcounter{equation}{0}
We have developed here the $rs$-matrix scheme for
natural finite-dimensional integrable systems  connected to the KdV
and the coupled KdV hierarchies. But it applies to other hierarchies
of integrable nonlinear evolution equations as well.
For the hierarchy of the KdV type equations the Lax matrix
 $M_0$ (\ref{vam}) is independent
of the spectral parameter. The corresponding matrices for
 the AKNS hierarchy
 and for the SG hierarchy have one pole at infinity or two poles at
infinity and at zero correspondingly  \cite{ft87}.
For applying our scheme in these cases one has to redefine
the projector $\P_\l$:
\bq
\P_{\l\m} =\v(\m)\cdot {\rm Res}^K_{\l=\infty}\,+\,
{\rm Res}^{K-1}_{\l=\infty}\,,\qquad
\v(\m)=\m {~}{~}{\rm or}{~}{~}\v(\m)=\m+\dfrac{1}{\m}\,.
\label{akns}
\eq
These two choices of the function $\v(\m)$ correspond to the
AKNS and to the SG hierarchy if we use the second rational
 parametrization for the $r$-matrix of $XXZ$ type (\ref{rxxz2})
 for the SG equation. The Hamiltonians in  terms of the root
variables $u_k$ of finite-dimensional systems connected with
stationary flows for this hierarchies were introduced
 in \cite{al81,almar92}.

In order to apply of our approach to restricted flows of the
 AKNS hierarchy of equations \cite{rrw92}
 we have to fix the form of the matrix $M_0$ as:
$$ M_0(\m)=\left(\begin{array}{cc}\m&w\\u&-\m\end{array}\right)\,,$$
and to choose the projector $\P_{\l\m}$ in the form
(\ref{akns}).

In the next step one has to consider the undeformed matrix $L_0(\l)$
 corresponding to this choice of $M_0$ and $\P_{\l\m}$ and to
prove that it obeys the $r$-matrix algebra. Let the
 matrix $L_0(\l)$
 has the form
$$L_0(\l)=L_0^\infty(\l)+L^{\sl D}_0(\l)$$
where we split $L_0(\l)$  into  the polynomial (pole at infinity)
 part and the $\sl D$-divisor part corresponding to  finite poles.
One finds the following sequence of $L^\infty_0$ matrices yielding
 the prescribed
 $M_0$ and $\P_{\l\m}$ and satisfying the $r$-matrix
 algebra (\ref{rpoi}):
\ben
L^\infty_0(\l)&=&\left(\begin{array}{cc}1&0\\0&-1\end{array}
\right)\,,\nn\\
L^\infty_0(\l)&=&\left(\begin{array}{cc}\l&q_1\\p_1&
-\l\end{array}\right)\,,\quad\{p_1,q_1\}=1\,,\nn\\
L^\infty_0(\l)&=&\left(\begin{array}{cc}\l^2-
\dfrac{q_1q_2}{2}&\l q_1-2p_2\\ \l
q_2-2p_1&-\l^2+\dfrac{q_1q_2}{2}\end{array}\right)\,,
\quad \{p_2,q_2\}=1\,.\nn
\en
The analog of the parametrization related to the  Newton
 representation (\ref{nrep}) \cite{rw92} for the recurrent
 construction of these matrices is not known as yet.
In this case the Hamiltonians are equal to $H=\P_{\l\m}\,[d(\l)]$,
 but they do not depend on the spectral parameter $\m$ since
 the corresponding residues are equal to zero.

The matrix $M_0$ is more symmetric by its entries in this case and
 respective integrable systems are of non-natural type \cite{rrw92}
 one can use various solutions of the equations
(\ref{alphs}) for the deformation coefficients $\a_j$.
Systems with such deformations will be investigated in details
in the forthcoming publications.

Finally we remark that  the $rs$-matrix scheme presented in this paper
applies to the discrete time analogs of soliton hierarchies.
 It yields an $rs$-matrix description of integrable symplectic maps.

\section{Acknowledgements}
 One of us (AVT) is very grateful to
 Swedish Royal Academy of Science
 NFR grant F-AA/MA 08677-326 for support
of a collaborative programm and Linkoping
University for warm hospitality.


\begin{thebibliography}{1}

\bibitem{avm80}
Adler,  M.,  van Moerbeke,  P.:
\newblock Completely integrable systems, {K}ac-{M}oody {Lie}
algebras and
  curves.
\newblock  Adv. Math {\bf 38}, 267--317 (1980)

\bibitem{avm94}
Adler,  M.,  van Moerbeke,  P.:
\newblock Compatible {P}oisson structure and the {Virasoro}
algebra.
\newblock Comm. Pure Appl. Math. {\bf 47}, 5--37 (1994)

\bibitem{al81}
Alber,  S.J.:
\newblock On stationary problem for equations of {K}orteweg-de
{V}ries type.
\newblock Comm. Pure Appl. Math. {\bf 34}, 259--272 (1981)

\bibitem{al92}
Alber,  M.S.:
\newblock Complex geometric asymptotics,  geometric
 phases and nonlinear
  integrable problems.
\newblock In: Ferwerda,  H.A.,  Blok,  H. and  Kuiken,
 H.K. (eds.)  Huygens'
  {P}rinciple 1690-1990,  {T}heory and {A}pplications,
 pp. 415--427. Elsevier
  Science Publishers B.V. 1992

\bibitem{almar92}
 Alber,  M.S.,   Marsden,  E.J.:
\newblock On geometric phases for soliton equations.
\newblock Commun. Math. Phys. {\bf 149}, 217--240 (1992)

\bibitem{arw92a}
Antonowicz,  M.,  Rauch-Wojciechowski,  S.:
\newblock Bi-hamiltonian formulation of the {H}\'enon-{H}eiles
 system and its
  multi-dimensional extensions.
\newblock Phys.Lett. {\bf A136}, 167--172 (1992)

\bibitem{arw92}
Antonowicz,  M.,  Rauch-Wojciechowski,  S.:
\newblock {H}ow to construct finite-dimensional bi-{H}amiltonian
 systems from
  soliton equations,  {J}acobi integrable potentials.
\newblock J. Math.Phys. {\bf 33}, 2115--2125 (1992)

\bibitem{arw93}
Antonowicz,  M.,  Rauch-Wojciechowski,  S.:
\newblock Soliton hierarchies with sources and {L}ax
representation for restricted flows.
\newblock Inv.Prob. {\bf 9}, 201--215 (1993)

\bibitem{bv90}
Babelon. O.,   Viallet,  C.M.:
\newblock Hamiltonian structures and {L}ax equations.
\newblock Phys. Lett. {\bf B237}, 411--416 (1990)

\bibitem{dnm76}
Dubrovin, B.A., Novikov, S.P. and Matveev, V.B.
\newblock Nonlinear equations of the {K}orteweg-de
{V}ries type, finite-band
  linear operators and abelian varietes.
\newblock Russ. Math. Surv., 31, 59--146, (1976)


\bibitem{eekt94}
Eilbeck,  J.C.,   Enolskii,  V.Z.,   Kuznetsov,  V.B.,
 Tsiganov,  A.V.:
\newblock {L}inear $r$-matrix algebra for classical
 separable systems.
\newblock  J.Phys. {\bf A27}, 567--578 (1994)

\bibitem{ft87}
Faddeev,  L.D.,  Takhtajan,  L.A.:
\newblock  Hamiltonain methods in the theory of solitons.
\newblock {S}pringer,  {B}erlin 1987

\bibitem{fld94}
 Faddeev,  L.D.:
\newblock Algebraic aspects of {B}ethe ansatz.
\newblock  Preprint,  Stony Brook,  ITP-SB-94-11,
 42pp. March 1994
\newblock HEP-TH/9404013.

\bibitem{fo91}
 Fordy,  A.P.:
\newblock The {H}\'enon-{H}eiles system revisited.
\newblock  Physica  {\bf D52}, 204--210 (1991)

\bibitem{rav93}
Gavrilov,  L.,  Ravozon,  V.,  Coboz,  R.:
\newblock {S}eparability and {L}ax pair for
 {H}enon-{H}eiles system.
\newblock  J.Math.Phys. {\bf 34}, 2385--2393 (1993)

\bibitem{kkm94}
Kalnins,  E.G.,  Kuznetsov,  V.B.,  Miller,  Jr. W.:
\newblock Quadrics on complex {R}iemannian spaces of
 constant curvature
  separation of variables and the {G}audin magnet.
\newblock J.Math.Phys. {\bf 35}, 1710--1731 (1994)


\bibitem{ks82}
 Kulish,  P.P.,  Sklyanin,  E.K.:
\newblock Integrable {Q}uantum {F}ield {T}heories.
\newblock In: Hietarinta,  J.,  Montonen,  C. (eds.)
  Lecture Notes in
  Physics {\bf 151} pp. 61--119,  Berlin Springer. 1982

\bibitem{krw94}
Kulish. P.P.,   Rauch-Wojciechowski,  S.,  Tsiganov,  A.V.:
\newblock {R}estricted flows of the {K}d{V} hierarchy
 and $r$-matrix formalism.
\newblock Mod.Phys.Lett. {\bf A9}, 2063--2073 (1994)

\bibitem{kuz92a}
Kuznetsov,  V.B.:
\newblock Equivalence of two graphical calculi.
\newblock  J. Phys. {\bf A25}, 6005--6025 (1992)


\bibitem{pe91}
Perelomov,  A.M.:
\newblock  Integrable system of classical mechanics
 and Lie algebras.
\newblock Birkha\"user,  Basel 1991

\bibitem{rrw92}
Ragnisco,  O.,  Rauch-Wojciechowski,  S.:
\newblock Restricted flows of the AKNS hierarchy.
\newblock Inverse Problems {\bf 8}, 245--262 (1992)

\bibitem{rav94}
Ramani,  A.,  Ravozon,  V.,  Grammaticos,  R.:
\newblock Generalized separability for a {H}amiltonian
 with nonseparable
  quartic potential.
\newblock  Phys.Lett. {\bf A191}, 91--95 (1994)
\bibitem{wo85}
 Rauch-Wojciechowski,  S.:
\newblock Three families of integrable one-particle
 potentials.
\newblock Preprint Inst. di Fisica,  Univ. di Roma
 La Sapienza 1985.

\bibitem{rw92}
 Rauch-Wojciechowski,  S.:
\newblock Newton representation for stationary flows
 of the {KdV} hierarchy.
\newblock Phys.Lett. {\bf A170}, 91--94 (1988)


\bibitem{rs87}
 Reyman,  A.G.,  and Semenov-{T}ian-{S}hansky,  M.A.:
\newblock Group {T}heoretical {M}ethods in the theory
 of {F}inite-{D}imensional
  {I}ntegrable systems.
\newblock In: Novikov,  S.P.,  Arnold,  V.I. (eds.),
 {D}ynamical systems {VII},
  EMS {\bf 16},  1986 Berlin,  Springer.1993

\bibitem{rs88}
Reyman,  A.G.,  and Semenov-{T}ian-{S}hansky,  M.A.:
\newblock Compatible {P}oisson structures for {Lax}
equations  an $r$-matrix
  approach.
\newblock Phys.Lett. {\bf A130}, 456--480 (1988)

\bibitem{skl86}
Sklyanin,  E.K.:
\newblock Poisson structure of the periodic classical
 {$XYZ$} chain.
\newblock  J.Soviet.Math. {\bf 46}, 1664--1683 (1989)

\bibitem{skl89}
Sklyanin,  E.K.:
\newblock Separation of variables in the {G}audin model.
\newblock J. Soviet. Math. {\bf 47}, 2473--2488 (1989)

\bibitem{skl92}
Sklyanin,  E.K.:
\newblock Separation of variables in the classical
integrable {$SL(3)$}
  magnetic chain.
\newblock Commun. Math. Phys. {\bf 142}, 123--132 (1992)

\bibitem{skl93}
Sklyanin,  E.K.:
\newblock Dynamical $r$-matrices for the elliptic
{C}alodgero-{M}ozer model.
\newblock Algebra i Anal. {\bf 6}, 227--238
(1993) (in Russian)

\bibitem{sur93a}
 Suris,  Yu.B.:
\newblock On the bi-{H}amiltonian structure of
{T}oda and relativistic {T}oda
  lattices.
\newblock  Phys.Lett. {\bf A180}, 419--433 (1993)

\bibitem{ts94}
Tsiganov,  A.V.:
\newblock Additive deformations of the {R}-matrix algebras.
\newblock J.Phys. {\bf A27}, 6759--6780 (1994)

\bibitem{van92}
Vanhaecke,  P.:
\newblock Linearising two dimensional integrable
systems and the construction
  of action-angle variables.
\newblock  Math.Z. {\bf 211}, 265--313 (1992)

\bibitem{w08}
Weber, H.:
\newblock {E}lliptische {F}unktionen und {A}lgebraische.
\newblock In: {V}ieweg, F. and {S}ohn (eds.)
Lehrbuch der {A}lgebra.
  {\bf {B}111},  Braunschweig Zahlen. 1908


\end{thebibliography}
\end{document}